\documentclass[a4paper,twocolumn,10pt,accepted=2021-07-15]{quantumarticle}
\pdfoutput=1
\usepackage{graphicx,tikz}
\input{tikzit.sty}
\tiny
\tikzstyle{rectangle}=[fill=white, draw=black, shape=rectangle]
\tikzstyle{red}=[fill=red, draw=black, shape=circle]
\tikzstyle{system_label}=[fill=white, draw=white, shape=circle]
\tikzstyle{prep}=[fill=white, draw=black, shape=rectangle, tikzit shape=rectangle, new atom]
\tikzstyle{four wire box}=[fill=white, draw=black, shape=rectangle, minimum width=0.70 cm, minimum height=3  cm]

\tikzstyle{lambda}=[-, draw={rgb,255: red,191; green,191; blue,191}]
\tikzstyle{state}=[<-]
\tikzstyle{new edge style 0}=[-]
\tikzstyle{new edge style 1}=[-]

\usepackage{xcolor}
\usepackage{amsmath,amssymb,mathtools,amsthm,mathrsfs,enumitem}
\usepackage[linkcolor=red]{hyperref}
\usepackage[matrix,frame,arrow]{xypic}

\usepackage[matrix,frame,arrow]{xy}
\usepackage{amsmath}

\newcommand{\qw}[1][-1]{\ar @{-} [0,#1]}

\newcommand{\gate}[1]{*{\xy *+<.6em>{#1};p\save+LU;+RU **\dir{-}\restore\save+RU;+RD **\dir{-}\restore\save+RD;+LD **\dir{-}\restore\POS+LD;+LU **\dir{-}\endxy} \qw}

\newcommand{\measureD}[1]{*{\xy*+=+<.5em>{\vphantom{\rule{0em}{.1em}#1}}*\cir{r_l};p\save*!R{#1} \restore\save+UC;+UC-<.5em,0em>*!R{\hphantom{#1}}+L **\dir{-} \restore\save+DC;+DC-<.5em,0em>*!R{\hphantom{#1}}+L **\dir{-} \restore\POS+UC-<.5em,0em>*!R{\hphantom{#1}}+L;+DC-<.5em,0em>*!R{\hphantom{#1}}+L **\dir{-} \endxy} \qw}

\newcommand{\multigate}[2]{*+<1em,.9em>{\hphantom{#2}} \qw \POS[0,0].[#1,0];p !C *{#2},p \save+LU;+RU **\dir{-}\restore\save+RU;+RD **\dir{-}\restore\save+RD;+LD **\dir{-}\restore\save+LD;+LU **\dir{-}\restore}

\newcommand{\ghost}[1]{*+<1em,.9em>{\hphantom{#1}} \qw}

\newcommand{\ustick}[1]{*!D!<0em,-.2em>=<0em>{{\scriptstyle #1}}}

\newcommand{\Qcircuit}[1][0em]{\xymatrix @*=<#1>} 
\newcommand{\node}[2][]{{\begin{array}{c} \ _{#1}\  \\ {#2} \\ \ \end{array}}\drop\frm{o} }

\newcommand{\pureghost}[1]{*+<1em,.9em>{\hphantom{#1}}}
\newcommand{\multiprepareC}[2]{*+<1em,.9em>{\hphantom{#2}}\save[0,0].[#1,0];p\save !C
  *{#2},p+RU+<0em,0em>;+LU+<+.8em,0em> **\dir{-}\restore\save +RD;+RU **\dir{-}\restore\save
  +RD;+LD+<.8em,0em> **\dir{-} \restore\save +LD+<0em,.8em>;+LU-<0em,.8em> **\dir{-} \restore \POS
  !UL*!UL{\cir<.9em>{u_r}};!DL*!DL{\cir<.9em>{l_u}}\restore}
\newcommand{\prepareC}[1]{*{\xy*+=+<.5em>{\vphantom{#1\rule{0em}{.1em}}}*\cir{l^r};p\save*!L{#1} \restore\save+UC;+UC+<.5em,0em>*!L{\hphantom{#1}}+R **\dir{-} \restore\save+DC;+DC+<.5em,0em>*!L{\hphantom{#1}}+R **\dir{-} \restore\POS+UC+<.5em,0em>*!L{\hphantom{#1}}+R;+DC+<.5em,0em>*!L{\hphantom{#1}}+R **\dir{-} \endxy}}

\newtheorem{lemma}{Lemma} 
 \newtheorem{theorem}{Theorem}
 \newtheorem{definition}{Definition}

\def\>{\rangle}
\def\<{\langle}

\def\trnsfrm#1{\mathscr #1}
 
	\def\rA{{\rm A}}
	\def\rB{{\rm B}}
	\def\rC{{\rm C}}
	\def\rD{{\rm D}} 
	\def\rE{{\rm E}} 
	\def\rF{{\rm F}}

    \def\rI{{\rm I}}  
     \def\rX{{\rm X}} 
    
\def\camp#1{\mathsf{#1}}
    \def\cX{\camp X}
    \def\cY{\camp Y}
    \def\cZ{\camp Z}
    \def\cW{\camp W}

\def\tA{\trnsfrm A}
\def\tB{\trnsfrm B}
\def\tC{\trnsfrm C}
\def\tD{\trnsfrm D}
\def\tG{\trnsfrm G}

\def\tK{\trnsfrm K}
\def\tI{\trnsfrm I}
\def\tT{\trnsfrm T}
\def\tU{\trnsfrm U}
\def\tV{\trnsfrm V} 
\def\tW{\trnsfrm W}

\def\tR{\trnsfrm R}
   
\def\tS{\trnsfrm S}
\def\Elety{\mathsf{Sys}}
\def\Testun{\mathsf{Test}}
\def\Evun{\mathsf{Ev}}
\def\test#1{\mathsf{#1}}
\def\teA{\test{A}}
\def\teB{\test{B}}
\def\teD{\test{D}}
\def\teE{\test{E}}

\def\teR{\test{R}}
\def\teS{\test{S}}
\def\teT{\test{T}} 
\def\teI{\test{I}}
\def\teW{\test{W}} 
\def\crng#1{\mathrm{#1}}
\def\mm{\crng m}
\def\mn{\crng n}
\def\ml{\crng l}

\def\Cntset#1{[\![\bar#1]\!]}

\def\aTset#1{[\![ #1]\!]} 
\def\Trnset#1{[\![#1]\!]}
\def\Testset#1{\<\!\<#1\>\!\>}

\def\Reals{{\mathbb R}}

\def\set#1{\mathsf{#1}}

\def\syst{\set{Sys}}
\def\TUAvar#1#2{\ensuremath{\tT_{#2}({#1})}}
\def\TUA{\TUAvar{\tU}{\rA}}
\def\tTUAvar#1#2{\ensuremath{\tilde\tT_{#2}({#1})}}
\def\tTUA{\tTUAvar{\tU}{\rA}}

\begin{document}

\title{Causal influence in operational probabilistic theories}
\author{Paolo Perinotti}\email{paolo.perinotti@unipv.it} 
\affiliation{{\em QUIT Group}, Dipartimento di Fisica, Universit\`a degli studi di Pavia, and INFN sezione di Pavia, via Bassi 6, 27100 Pavia, Italy}
\homepage{http://www.qubit.it}
\email{paolo.perinotti@unipv.it}
\orcid{0000-0003-4825-4264}

\maketitle

\begin{abstract}
We study the relation of  causal influence between input systems of a reversible evolution and its output 
systems, in the context of operational probabilistic theories. We analyse two different definitions that are 
borrowed from the literature on quantum theory---where they are equivalent. One is the notion based on 
signalling, and the other one is the notion used to define the neighbourhood of a cell in a quantum cellular 
automaton. The latter definition, that we adopt in the general scenario, turns out to be strictly weaker than the 
former: it is possible for a system to have causal influence on another one without signalling to it. Remarkably, 
the counterexample comes from classical theory, where the proposed notion of causal influence determines a 
redefinition of the neighbourhood of a cell in cellular automata. We stress that, according to our definition, it 
is impossible anyway to have causal influence in the absence of an interaction, e.g.~in a Bell-like scenario.
We study various conditions for causal influence, and introduce the feature that we call 
{\em no interaction without disturbance}, under which we prove that signalling and causal influence coincide. 
The proposed definition has interesting consequences on the analysis of causal networks, and leads
to a revision of the notion of neighbourhood for classical cellular automata, clarifying a puzzle regarding their 
quantisation that apparently makes the neighbourhood larger than the original one.
\end{abstract}
\maketitle

\section{Introduction}

In the last two decades, studies on the foundations of Quantum Theory flourished, nurtured by the wealth of results in quantum information 
theory and their impact on the understanding of the quantum realm. One of the many facets of quantum theory that were explored in this perspective
is causal influence and the consequent analysis of causal 
structures~ 
\cite{Beckman:2001aa,Eggeling_2002,Schumacher:2005aa,Pawowski:2009aa,Chiribella_2008,PhysRevA.88.052130,Milz2020kolmogorovextension,Giarmatzi:2018aa,Cotler:2019aa}. 
In this line of thought, the main questions regard possible criteria to determine the compatibility of observed data regarding a network 
of systems and hypothetical relations between them. The techniques adopted for this kind of analysis were initially developed in the context of
classical Bayesian networks~\cite{pearl_2009}, or in the theory of quantum networks developed by different 
approaches~\cite{PhysRevA.88.052130,10.1145/1250790.1250873,PhysRevA.80.022339,PhysRevA.88.022318,Ried:2015aa,barrett2019quantum,Oreshkov2012aa}.  A second interesting question analysed in the literature regards quantification of causal influence~\cite{Cotler:2019aa}.
However, the need for abstracting the study of causal structures from the specific formalism of the theory adopted was pointed out already in Ref.~\cite{Hardy_2007}, where the purpose was to build a microscopic theory of gravity that might require a step beyond quantum. In this perspective, a language that allows one to deal with general information processing structures without the burden of using a specific mathematical model is precisely the right tool.

The framework of Operational Probabilistic Theories (OPTs)~\cite{PhysRevA.81.062348,PhysRevA.84.012311,DAriano:2017aa}, is a landscape of 
theories each of which could in principle provide a mathematical language for representing elementary systems and their processes, different form the 
classical or quantum one, but bearing important common features with those ones. OPTs are indeed defined as all the possible theories that share with 
classical or quantum theory some basic structure: in particular, the properties of rules by which one can form composite  processes as sequences of 
other processes, form composite systems, or apply processes independently on subsystems of a composite system, as well as the properties of rules to 
calculate probabilities of events that can occur as alternative outcomes within a process. The mentioned features are sufficient to prove results 
that hold for all theories, or for wide classes of them, characterised by some operational property, 
e.g.~purification~\cite{PhysRevA.81.062348,DAriano:2017aa}, or $n$-local discriminabilty~\cite{Hardy:2012to}, and so on. Understanding entanglement~\cite{Chiribella_2015} and its relation with complementarity~\cite{PhysRevA.101.042118,PhysRevA.102.052216}, information and disturbance~\cite{DAriano2020information}, as well as any other fundamental feature within a general OPT allows us to 
understand the feature itself beyond the contingent aspects that it acquires within a specific theory.

The language of OPTs is then perfectly suited to the purpose of analysing causal relations and causal structures beyond the classical or quantum 
realm. In the present study, we consider the relation of causal influence between systems within the OPT framework. 
We provide here a definition of causal influence in reference to a reversible evolution, based on 
how the evolution propagates the effects of an 
intervention, but we do not identify the propagated effect with the ability to signal, i.e.~to transmit 
information. We then discuss the consequences of our definition, and provide a necessary and sufficient condition 
for its fulfilment. We study the relation between the proposed definition and the property of signalling, and 
show that in general our definition is strictly weaker than the traditional one based on signalling, i.e.~one
system can have causal influence on another one without signalling to it. An example of a theory where our 
definition is strictly weaker than signalling is classical information theory. This fact bears important 
consequences. As a remarkable example, think of the neighbourhood of a cell $C$ in a cellular automaton: the 
neighbourhood of $C$ is usually defined as the set of cells to which $C$ can signal in one step. However, if we  
define the neighbourhood of $C$ as the set of cells on which $C$ can have causal influence in one step instead, 
the latter neighbourhood will be generally larger than the former one. Remarkably, some structural theorems 
for quantum cellular automata fail in the classical case, but keep holding if one redefines the 
neighbourhood according to the definition presented here (see e.g.~\cite{Perinotti2020cellularautomatain})

We then study conditions under which the two definitions coincide.
For this purpose, we introduce the property of {\em no interaction without disturbance}, which essentially means 
that every non trivial interaction with a system enables signalling to it. As one can intuitively expect, in 
a theory with no interaction without disturbance---e.g.~Quantum Theory or Fermionic Theory---causal 
influence and signalling coincide. Finally we construct a useful tool for the analysis of the causal structure of 
a given process. Remarkably, the latter turns out to be very powerful in detecting causal influence 
relations, to the extent that the case of non-local activation of causal influence is excluded, and the analysis 
of the causal neighbourhood of a system can be carried out locally.

In everyday experience, in order to establish a causal influence relation between two systems one has to perform 
controlled interventions on one system and then look for consequences on the other one, after a given time 
interval. As an example, consider a point $x_0$ in space. If we adopt an inertial reference frame and 
activate an electromagnetic source at $x_0$ at time $t_0$, assuming an isotropic medium, we expect a spherical 
wave front centred at $x_0$ to start propagating isotropically at the speed of light. One can then detect a pulse 
in any direction, e.g. at point $x_1$ after a time given by the radius of the sphere centred at $x_0$ and 
containing $x_1$, divided by the speed of light in the medium. In this case we claim that the perturbation of the 
electromagnetic field detected at $x_1$ is a consequence of the initial intervention at $x_0$. This way of 
revealing a causal influence relation between two systems mixes two distinct notions: the first one is based on 
testing the consequences of a controlled intervention occurred on the first 
system as they propagate to the second system, and the second one is based on identifying those consequences as 
the possibility of transmitting information. Here we decouple these two aspects, and focus on the first one.

The question underpinning our definition is the following: if we had to simulate the consequences of a 
hypothetical intervention on system $\rA$ occurred {\em before} the evolution $\tU$, by intervening {\em after} 
$\tU$ instead, would a non-trivial action on system $\rB$ be required? If the answer is positive, we will 
say that the evolution $\tU$ mediates a causal influence from $\rA$ to $\rB$.


\section{Operational probabilistic theories}\label{sec:opts}

In the present section we provide a brief review of the framework of operational probabilistic theories, with focus on those definitions that
will be used in the remainder. For a comprehensive account of the subject, the reader is referred 
to~\cite{PhysRevA.81.062348,DAriano:2017aa,Chiribella2016,Perinotti2020cellularautomatain}. 

%

An operational probabilistic theory provides a language to describe processes occurring on systems along with rules for the composition of processes
in parallel or in a sequence, and other rules for the calculation of probabilities of events that might occur in a network of processes. The first set 
of rules defines an operational theory $\Theta$, which consists in the following elements.
\begin{enumerate}
\item A collection $\Testun(\Theta)$ of tests. A test will be denoted by 
$\teT^{\rA\to\rB}_\cX$, where the label $\rA\to\rB$ emphasises the type of input and output system for the test, and will be omitted when the latter 
are manifest from the context, while $\cX$ represents the finite set of outcomes corresponding to the individual events that might occur within the 
test. The collection of system types will be denoted by $\Elety(\Theta)$. The set of tests of type $\rA\to\rB$ is denoted as $\Testset{\rA\to\rB}$ 
\item An associative rule for sequential composition: the test 
$\teT_\cX\in\Testset{\rA\to\rB}$ can be followed by the test $\teS_\cY\in\Testset{\rB\to\rC}$, thus obtaining the sequential composition 
$\teS\teT_{\cX\times\cY}\in\Testset{\rA\to\rC}$. 
\item A rule $\otimes:(\rA,\rB)\mapsto\rA\rB$ 
for composing labels in parallel, with the following properties
\begin{enumerate}
\item Associativity: $(\rA\rB)\rC=\rA(\rB\rC)$. 
\item Unit: there is a label $\rI$ such that $\rI\rA=\rA\rI=\rA$ for every $\rA\in\Elety(\Theta)$.
\end{enumerate}
A corresponding rule for composition of tests $\otimes:\Testset{\rA\to\rB}\times\Testset{\rC\to\rD}\to\Testset{\rA\rC\to\rB\rD}$
such that $\otimes:(\teS_\cX,\teT_\cY)\mapsto(\teS\otimes\teT)_{\cX\times\cY}\in\Testset{\rA}$. Also this rule has its properties
\begin{enumerate}
\item Associativity:  
\begin{align*}
(\teS_\cX\otimes\teT_\cY)\otimes\teW_\cZ=\teS_\cX\otimes(\teT_\cY\otimes\teW_\cZ). 
\end{align*}
\item Identity: for every $\rA\in\Elety(\Theta)$, there is a test $\teI_\rA\in\Testset{\rA\to\rA}$ such that $\teI_\rB\teS_\cX=\teS_\cX\teI_\rA=\teS_\cX$, for every $\teS_\cX\in\Testset{\rA\to\rB}$.
\item For every $\teA_\cX\in\Testset{\rA\to\rB}$, $\teB_\cY\in\Testset{\rB\to\rC}$, $\teD_\cZ\in\Testset{\rD\to\rE}$, $\teE_\cW\in\Testset{\rE\to\rF}$, one has 
\begin{align}
(\teB_\cY\otimes\teE_\cW)(\teA_\cX\otimes\teD_\cZ)=(\teB_\cY\teA_\cX)\otimes(\teE_\cW\teD_\cZ).
\label{eq:natur}
\end{align}
\item \label{it:braid}Braiding: for every pair of system types $\rA,\rB$, there exist tests $\teS_{\rA\rB},\teS^*_{\rA\rB}\in\Testset{\rA\rB\to\rB\rA}$ such that $\teS^*_{\rB\rA}\teS_{\rA\rB}=\teS_{\rB\rA}\teS^*_{\rA\rB}=\teI_{\rA\rB}$, and $\teS_{\rA\rB}(\teA_\cX\otimes\teB_\cY)=(\teB_\cY\otimes\teA_\cX)\teS_{\rA\rB}$. Moreover, 
\begin{align*}
&\teS_{(\rA\rB)\rC}=(\teI_\rA\otimes\teS_{\rB\rC})(\teS_{\rA\rC}\otimes\teI_\rB),\\
&\teS_{\rA(\rB\rC)}=(\teS_{\rA\rB}\otimes\teI_\rC)(\teI_\rB\otimes\teS_{\rA\rC}).
\end{align*}
When $\teS^*_{\rA\rB}\equiv\teS_{\rA\rB}$, the theory is {\em symmetric}. 
\end{enumerate}
All the theories developed so far are symmetric. 
\end{enumerate}

All tests of an operational theory are (finite) collections of {\em events}: $\Testset{\rA\to\rB}\ni\teR_\cX=\{\tR_i\}_{i\in\cX}$. If 
$\teR_\cX\in\Testset{\rA\to\rB}$ and $\teT_\cY\in\Testset{\rB\to\rC}$, then 
\begin{align*}
\Testset{\rA\to\rC}\ni(\teT\teR)_{\cX\times\cY}\coloneqq \{\tT_j\tR_i\}_{(i,j)\in\cX\times\cY}. 
\end{align*}
Similarly, for $\teR_\cX\in\Testset{\rA\to\rB}$ and $\teT_\cY\in\Testset{\rC\to\rD}$,
\begin{align*}
(\teR\otimes\teT)_{\cX\times\cY}\coloneqq \{\tR_i\otimes\tT_j\}_{(i,j)\in\cX\times\cY}.
\end{align*}
The set of all events of all tests in $\Testset{\rA\to\rB}$ is denoted by $\Trnset{\rA\to\rB}$.
By the properties of sequential and parallel composition of tests, one can easily derive associativity of sequential and parallel composition of events, as well as the analogue of Eq.~\eqref{eq:natur}. For every test $\teT_\cX\in\Testset{\rA\to\rB}$ with $\teT_\cX=\{\tT_i\}_{i\in\cX}$, and every disjoint partition $\{\cX_j\}_{j\in\cY}$ of $\cX=\bigcup_{j\in\cY}\cX_j$, one has a {\em coarse graining} operation that maps $\teT_\cX$ to $\teT'_\cY\in\Testset{\rA\to\rB}$, with $\teT'_\cY=\{\tT'_j\}_{j\in\cY}$. We define $\tT_{\cX_j}\coloneqq \tT'_j$. If an element of the partition is trivial, i.e.~$\cX_j=\{i_0\}$, then $\tT_{\cX_j}=\tT_{i_0}$. The parallel and sequential compositions distribute over coarse graining: 
\begin{align*}
&\tT_{\cX_j}\otimes\tR_{\cY_k}=(\tT\otimes\tR)_{\cX_j\times \cY_k},\\
&\tA_{\cY_l}\tT_{\cX_j}\tB_{\cZ_k}=(\tA\tT\tB)_{\cY_l\times\cX_j\times \cZ_k}.
\end{align*}
Notice that for every test $\teT_\cX\in\Testset{\rA\to\rB}$ there exists the singleton test 
$\teT'_*\coloneqq \{\tT_\cX\}$. One can easily prove that the identity test $\teI_{\rA}$ is a singleton: 
$\teI_{\rA}=\{\tI_{\rA}\}$, and $\tI_\rB\tT=\tT\tI_\rA$ for every event $\tT\in\Trnset{\rA\to\rB}$. Similarly, 
for $\teS_{\rA\rB}=\{\tS_{\rA\rB}\}$ and $\teS^*_{\rA\rB}=\{\tS^*_{\rA\rB}\}$ we have 
$\tS^*_{\rB\rA}\tS_{\rA\rB}=\tS_{\rB\rA}\tS^*_{\rA\rB}=\tI_{\rA\rB}$. The unique event in a 
singleton test is called a \emph{channel}, and the collection of channels with input $\rA$ and output $\rB$ 
deserves its own symbol: $\Trnset{\rA\to\rB}_1$. In the special cases where $\rA=\rI$ or $\rB=\rI$, we have
the sets $\aTset{\rB}_1$ and $\Cntset{\rA}_1$, of \emph{deterministic} states and effects, respectively.

If the event $\tA_{\rX_j}$ belongs to a test that is obtained by 
coarse graining, for every $i\in\rX_j$ we will say that $\tA_i$ \emph{refines} $\tA_j$.

The collection of events of an operational theory $\Theta$ will be denoted by $\Evun(\Theta)$. The above 
requirements make the collections $\Testun(\Theta)$ and $\Evun(\Theta)$ the families of morphisms of two braided 
strict monoidal categories with the same objects---system types $\Elety(\Theta)$.

An operational theory is an OPT if the tests $\Testset{\rI\to\rI}$ are probability distributions: $1\geq\tT_i=p_i\geq0$, so that $\sum_{i\in\cX}p_i=1$, and given two tests $\teS_\cX,\teT_\cY\in\Testset{\rI\to\rI}$ with $\tS_i=p_i$ and $\tT_i=q_i$, the following identities hold
\begin{align*}
&\tS_i\otimes \tT_j=\tS_i\tT_j\coloneqq p_iq_j,\\
&\tT_{\cX_j}\coloneqq \sum_{i\in\cX_j}p_i,
\end{align*}
meaning that events in the same test are mutually exclusive and events in different tests 
of system $\rI$ are independent. While it is immediate that $1\in\Trnset{\rI\to\rI}$---since $\{1\}_*=\{\tI_\rI\}$ is the only singleton test---we will assume that 
$0\in\Trnset{\rI\to\rI}$. This means that we can 
consider e.g.~tests of the form $\{1,0,0\}$.

Events in $\Trnset{\rA\to\rB}$ are called {\em transformations}. 
As a consequence of the above definitions, 
every set $\aTset{\rA}\coloneqq \Trnset{\rI\to\rA}$ can be viewed as a set of functionals on 
$\Cntset{\rA}$. As such, it can be viewed as a spanning subset of the real vector space 
$\aTset{\rA}_\Reals$ of linear functionals on $\Cntset{\rA}$. On the other hand 
$\Cntset{\rA}\coloneqq \Trnset{\rA\to\rI}$ is a separating set of 
positive linear functionals on $\aTset{\rA}$, which then spans the dual space 
$\aTset{\rA}_{\Reals}^*=:\Cntset{\rA}_{\Reals}$. The dimension $D_\rA$ of $\aTset{\rA}_\Reals$ 
(which is the same as that of $\Cntset{\rA}_\Reals$) is called {\em size} of system $\rA$.  
Given a transformation $\tA\in\Trnset{\rA\to\rB}$ and a system $\rE$, $\tA\otimes\tI_\rE$ induces a 
linear map from $\aTset{\rA\rE}_\Reals$ to $\aTset{\rB\rE}_\Reals$. Every transformation can thus be identified 
by the infinite family of such linear maps that it induces. In an OPT, coarse graining is represented by the sum:
$\tA_{\rX_i}=\sum_{j\in\rX_i}\tA_j$. We will say that $\tA\in\Trnset{\rA\to\rB}$ is 
\emph{atomic} if every refinement of $\tA$ is trivial, i.e.~if $\tA_i=\sum_{j\in\rX_i}\tB_j$ implies that 
$\tB_j=p_j\tA_i$ for every $j\in\rX_i$.
Using the properties of parallel composition, one can prove that $D_{\rA\rB}\geq D_\rA D_\rB$. Events in 
$\aTset{\rA}$ are called {\em states}, and denoted by lower-case greek letters, e.g.~$\rho$, while events in 
$\Cntset{\rA}$ are called {\em effects}, and denoted by lower-case latin letters, e.g.~$a$. When it is 
appropriate, we will use the symbol $|\rho)$ to denote a state, and $(a|$ to denote an effect.
We will also use the circuit notation, where we denote states, transformations and effects by the symbols
\begin{align*}
\begin{aligned}
    \Qcircuit @C=1em @R=.7em @! R {&\prepareC{\rho}&\ustick{\rA}\qw&\qw}
\end{aligned}\ ,
\begin{aligned}
    \Qcircuit @C=1em @R=.7em @! R {&\ustick{\rA}\qw&\gate{\tA}&\ustick{\rB}\qw&\qw}
\end{aligned}\ ,
\begin{aligned}
    \Qcircuit @C=1em @R=.7em @! R {&\ustick{\rA}\qw&\measureD{a}}
\end{aligned}\ ,
\end{align*}
respectively. 
Sequential composition of $\tA\in\Trnset{\rA\to\rB}$  and $\tB\in\Trnset{\rB\to\rC}$ is 
denoted by the diagram
\begin{align*}
\begin{aligned}
    \Qcircuit @C=1em @R=.7em @! R {&\ustick{\rA}\qw&\gate{\tB\tA}&\ustick{\rC}\qw&\qw}
\end{aligned}\ =\ 
\begin{aligned}
    \Qcircuit @C=1em @R=.7em @! R {&\ustick{\rA}\qw&\gate{\tA}&\ustick{\rB}\qw&\gate{\tB}&\ustick{\rC}\qw&\qw}
\end{aligned}\ .
\end{align*}
For composite systems we use diagrams with multiple wires, e.g.
\begin{align*}
\begin{aligned}
    \Qcircuit @C=1em @R=.7em @! R {&\ustick{\rA}\qw&\multigate{1}{\tA}&\ustick{\rB}\qw&\qw\\
    &\ustick{\rC}\qw&\ghost{\tA}&\ustick{\rD}\qw&\qw}
\end{aligned}.
\end{align*}
The identity  will be omitted: $\Qcircuit @C=1em @R=.7em @! R {&\ustick{\rA}\qw&\gate{\tI}&\ustick{\rA}\qw&\qw}=    \Qcircuit @C=1em @R=.7em @! R {&\ustick{\rA}\qw&\qw}$. 
The swap $\tS_{\rA\rB}$ and its inverse will be denoted as follows
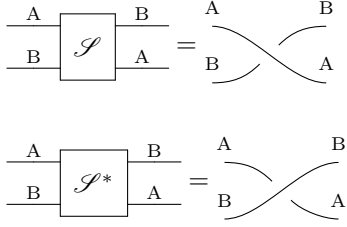
\begin{align*}
&
\begin{aligned}
    \Qcircuit @C=1em @R=.7em @! R {&\ustick{\rA}\qw&\multigate{1}{\tS}&\ustick{\rB}\qw&\qw\\
    &\ustick{\rB}\qw&\ghost{\tS}&\ustick{\rA}\qw&\qw}
\end{aligned}=
\begin{tikzpicture}
	\begin{pgfonlayer}{nodelayer}
		\node [style=none] (20) at (1.5, 1) {$\scriptstyle\rB$};
		\node [style=none] (19) at (-1.5, -0.5) {$\scriptstyle\rB$};
		\node [style=none] (0) at (-1.5, -1) {};
		\node [style=none] (1) at (-1.5, 0.5) {};
		\node [style=none] (3) at (1.5, -1) {};
		\node [style=none] (5) at (-1.5, -1) {};
		\node [style=none] (22) at (-0.25, -0.5) {};
		\node [style=none] (23) at (1.5, 0.5) {};
		\node [style=none] (24) at (0.25, 0) {};
		\node [style=none] (25) at (-1.5, 1) {$\scriptstyle\rA$};
		\node [style=none] (26) at (1.5, -0.5) {$\scriptstyle\rA$};
	\end{pgfonlayer}
	\begin{pgfonlayer}{edgelayer}
		\draw [in=-180, out=0, looseness=0.75] (1.center) to (3.center);
		\draw [in=-135, out=0] (5.center) to (22.center);
		\draw [in=45, out=-180] (23.center) to (24.center);
	\end{pgfonlayer}
\end{tikzpicture}\\
\\
&
\begin{aligned}
    \Qcircuit @C=1em @R=.7em @! R {&\ustick{\rA}\qw&\multigate{1}{\tS^*}&\ustick{\rB}\qw&\qw\\
    &\ustick{\rB}\qw&\ghost{\tS^*}&\ustick{\rA}\qw&\qw}
\end{aligned}=
\begin{tikzpicture}
	\begin{pgfonlayer}{nodelayer}
		\node [style=none] (20) at (1.5, 1) {$\scriptstyle\rB $};
		\node [style=none] (19) at (-1.5, -0.5) {$\scriptstyle\rB$};
		\node [style=none] (1) at (-1.5, -1) {};
		\node [style=none] (3) at (1.5, 0.5) {};
		\node [style=none] (5) at (-1.5, 0.5) {};
		\node [style=none] (22) at (-0.25, 0) {};
		\node [style=none] (23) at (1.5, -1) {};
		\node [style=none] (24) at (0.25, -0.5) {};
		\node [style=none] (25) at (-1.5, 1) {$\scriptstyle\rA$};
		\node [style=none] (26) at (1.5, -0.5) {$\scriptstyle\rA$};
	\end{pgfonlayer}
	\begin{pgfonlayer}{edgelayer}
		\draw [in=-180, out=0, looseness=0.75] (1.center) to (3.center);
		\draw [in=135, out=0] (5.center) to (22.center);
		\draw [in=-45, out=180, looseness=0.75] (23.center) to (24.center);
	\end{pgfonlayer}
\end{tikzpicture}
\end{align*}
In the present paper we will always assume that the theory under consideration is 
symmetric, however all the results can be straightforwardly generalised to the nontrivially braided case. We will
consequently draw the swap $\tS_{\rA\rB}$ as
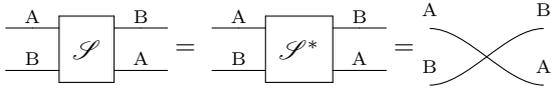
\begin{align*}
&
\begin{aligned}
    \Qcircuit @C=1em @R=.7em @! R {&\ustick{\rA}\qw&\multigate{1}{\tS}&\ustick{\rB}\qw&\qw\\
    &\ustick{\rB}\qw&\ghost{\tS}&\ustick{\rA}\qw&\qw}
\end{aligned}=\ 
\begin{aligned}
    \Qcircuit @C=1em @R=.7em @! R {&\ustick{\rA}\qw&\multigate{1}{\tS^*}&\ustick{\rB}\qw&\qw\\
    &\ustick{\rB}\qw&\ghost{\tS^*}&\ustick{\rA}\qw&\qw}
\end{aligned}=
\begin{tikzpicture}
	\begin{pgfonlayer}{nodelayer}
		\node [style=none] (20) at (1.5, 1) {$\scriptstyle\rB $};
		\node [style=none] (19) at (-1.5, -0.5) {$\scriptstyle\rB$};
		\node [style=none] (0) at (-1.5, -1) {};
		\node [style=none] (1) at (-1.5, 0.5) {};
		\node [style=none] (3) at (1.5, -1) {};
		\node [style=none] (5) at (-1.5, -1) {};
		\node [style=none] (22) at (1.5, 0.5) {};
		\node [style=none] (25) at (-1.5, 1) {$\scriptstyle\rA$};
		\node [style=none] (26) at (1.5, -0.5) {$\scriptstyle\rA$};
	\end{pgfonlayer}
	\begin{pgfonlayer}{edgelayer}
		\draw [in=-180, out=0, looseness=0.75] (1.center) to (3.center);
		\draw [in=-180, out=0, looseness=0.75] (5.center) to (22.center);
	\end{pgfonlayer}
\end{tikzpicture}
\end{align*}

An OPT $\Theta$ is specified by the collections of systems and tests, along with the parallel composition rule $\otimes$ 
\begin{align*}
\Theta\equiv(\Testun(\Theta),\Elety(\Theta),\otimes). 
\end{align*}

\section{Defining causal influence}\label{sewc:causinf}
Let us consider a channel $\tC\in\Trnset{\rA\rB\to\rA'\rB'}$. The kind of question we would like to address here is under what conditions 
are we allowed to say that, through the evolution given by $\tC$, interventions on system $\rA$ can influence system $\rB'$. In the first place, we 
remark that the question is meaningful without further discussion only within theories where no information is allowed to flow from the ``future'' 
towards the ``past''. As proved in Ref.~\cite{PhysRevA.81.062348}, the latter requirement is equivalent to uniqueness of the deterministic effect 
$e_\rA$ for every system $\rA$ of the theory. This property will then be assumed in the remainder.

Typically, the answer to this kind of question comes in the first place by stating when there is {\em no influence} from $\rA$ to $\rB'$, and then 
defining causal influence by negating the above condition. The no-influence relation normally boils down to the impossibility of using the channel 
$\tC$ to send a message from $\rA$ to $\rB'$---the so-called no-signalling condition. Considering the question in the quantum realm allowed various authors to highlight some aspects that 
warn us about the possible answers. In the first place, when $\tC$ is not reversible, one cannot trust conditions for no-influence. Indeed, 
considering that in quantum theory every channel can be viewed as the effective description of a reversible channel on an extended system, once we 
neglect the ``environment'', no-influence might be an accident due to the preparation of a special initial state of the environment rather than a 
structural feature of the channel $\tC$ itself. The above issue is strictly connected, in the quantum case, with a further one: it may happen that 
$\rA$ 
has no influence on $\rA'$ and on $\rB'$ individually, but has influence on the composite system $\rA'\rB'$. A typical example was exhibited in  
Ref.~\cite{PhysRevA.74.012305}, where the authors show a channel that allows for no influence from $\rA$ to either $\rA'$ or $\rB'$ and no influence  
from $\rB$ to either $\rA'$ or $\rB'$, but such that both $\rA$ and $\rB$ have influence on $\rA'\rB'$, and actually a reversible dilation of such  
channel highlights a finer structure: there are dilations where both $\rA$ and $\rB$ influence $\rA'$ and not $\rB'$, and others where they both influence $\rB'$ and not $\rA'$~\cite{PhysRevLett.106.010501}.

For the reasons listed above, we will define causal influence under {\em reversible} channels. For all those channels that can be obtained from 
reversible ones neglecting the environment as in the following scheme 
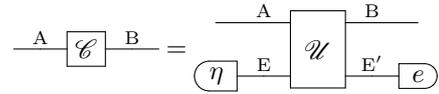
\begin{align*}
	\begin{aligned}
	\Qcircuit @C=1em @R=1em
	{
		&\ustick{\rA}\qw&\gate{\tC}&\ustick{\rB}\qw&\qw
	}
	\end{aligned}
	&=
	\begin{aligned}
	\Qcircuit @C=1em @R=1em
	{
		&\ustick{\rA}\qw&\multigate{1}{\tU}&\ustick{\rB}\qw&\qw
		\\
		\prepareC{\eta}&\ustick{\rE}\qw&\ghost{\tU}&\ustick{\rE'}\qw&\measureD{e}
	}
	\end{aligned}
\end{align*}
we will adopt only those influence relations compatible with each of its reversible 
dilations $\tU$. As discussed above, in quantum theory there are examples where influence relations depend on the particular dilation, and there is no influence relation that is actually independent of the reversible dilation $\tU$. 

It is probably useful to remind here that in all known theories {\em every} channel can be dilated to a 
reversible one, except for the theory of Popescu-Rohrlich boxes. The discussion of the definition of causal influence for irreversible channels
is beyond the purpose of the present study. 

We remark that outside the quantum, in particular in theories without local discriminability, the mentioned definition based on no-signalling can have 
a sort of ``dual'' issue to the one highlighted above: while neither $\rA$ nor $\rB$ signal to, say, $\rA'$, it may happen that the composite system 
$\rA\rB$ signals to $\rA'$.

Finally, we anticipate here that the definition that we will give is not the usual one based on the possibility or impossibility of signalling. 
Our definition represents a generalisation of the traditional one, and is strictly weaker: we will show examples of non-signalling channels that 
however allow for causal influence. In the special case of quantum theory the two definitions coincide.

\subsection{The definition}

We now give the definition of {\em no causal influence from $\rA$ to $\rB'$}. The definition differs from the 
usual one---given in terms of no-signalling conditions---and is taken from the literature on quantum cellular automata, where the neighbourhood of a 
given system is defined through the image of its local algebra of operators under conjugation by the unitary map representing the cellular 
automaton~\cite{schumacher2004reversible}. The definition for general OPTs, inspired by the latter, was given in 
Ref.~\cite{Perinotti2020cellularautomatain}.
\begin{definition}
Let $\tU\in\Trnset{\rA\rB\to\rA'\rB'}$ be reversible. We say that $\tU$ does not produce a causal influence of $\rA$ on $\rB'$, and write 
$\rA\not\rightarrow_\tU\rB'$, if for every $\rE$ and every $\tA\in\Trnset{\rA\rE\to\rA\rE}$ one has
\begin{align}
	&\begin{aligned}
	\Qcircuit @C=1em @R=1em
	{&\ustick{\rE}\qw&\qw&\qw&\multigate{1}{\tA}&\qw&\qw&\ustick{\rE}\qw&\qw\\
		&\ustick{\rA'}\qw&\multigate{1}{\tU^{-1}}&\ustick{\rA}\qw&\ghost{\tA}&\ustick{\rA}\qw&\multigate{1}{\tU}&\ustick{\rA'}\qw&\qw\\
		&\ustick{\rB'}\qw&\ghost{\tU^{-1}}&\qw&\ustick{\rB}\qw&\qw&\ghost{\tU}&\ustick{\rB'}\qw&\qw	}
	\end{aligned}\nonumber\\
	\nonumber\\
=&\ 
		\begin{aligned}
	\Qcircuit @C=1em @R=1em
	{	&\ustick{\rE}\qw&\multigate{1}{\tA'}&\ustick{\rE}\qw&\qw\\
		&\ustick{\rA'}\qw&\ghost{\tA'}&\ustick{\rA'}\qw&\qw\\
		&\qw&\ustick{\rB'}\qw&\qw&\qw
	}
	\end{aligned}\ .
\label{eq:noci}
\end{align}
for some $\tA'\in\Trnset{\rA'\rE\to\rA'\rE}$. On the contrary, we say that $\tU$ produces a causal influence of $\rA$ on $\rB'$, and write 
$\rA\rightarrow_\tU\rB'$, if there exists a system $\rE$ and $\tA\in\Trnset{\rA\rE\to\rA\rE}$ such that, for any $\tA'\in\Trnset{\rA'\rE\to\rA'\rE}$, condition~\eqref{eq:noci} 
does not hold.
\end{definition}

Before checking the consequences of the present definition and finding equivalent conditions, we will comment on its relation with other definitions in the literature. The definition based on no-signalling is the following
\begin{definition}
Let $\tU\in\Trnset{\rA\rB\to\rA'\rB'}$ be reversible. We say that $\tU$ does not allow for signalling from $\rA$ to $\rB'$, and write 
$\rA\not\leadsto_\tU\rB'$, if the following identity holds
\begin{align}
	\begin{aligned}
	\Qcircuit @C=1em @R=1em
	{	&\ustick{\rA}\qw&\multigate{1}{\tU}&\ustick{\rA'}\qw&\measureD{e}\\
		&\ustick{\rB}\qw&\ghost{\tU}&\ustick{\rB'}\qw&\qw	}
	\end{aligned}
	&\ =\ 
		\begin{aligned}
	\Qcircuit @C=1em @R=1em
	{	&\ustick{\rA}\qw&\measureD{e}\\
		&\ustick{\rB}\qw&\gate{\tC}&\ustick{\rB'}\qw&\qw
	}
	\end{aligned}\ .
\label{eq:nosig}
\end{align}
for some $\tC\in\Trnset{\rB\to\rB'}_1$. On the contrary, we say that $\tU$ allows for signalling from $\rA$ to $\rB'$, and write 
$\rA\leadsto_\tU\rB'$, if condition~\eqref{eq:nosig} does not hold.
\end{definition}
Now, we can easily show that if $\rA\not\rightarrow_\tU\rB'$, then $\rA\not\leadsto_\tU\rB'$. Indeed, let us first of all rewrite 
condition~\eqref{eq:noci} as follows: for every $\rE$ and every $\tA\in\Trnset{\rA\rE\to\rA\rE}$ one has
\begin{align}
	\begin{aligned}
	\Qcircuit @C=1em @R=1em
	{&\ustick{\rE}\qw&\multigate{1}{\tA}&\qw&\qw&\ustick{\rE}\qw&\qw\\
		&\ustick{\rA}\qw&\ghost{\tA}&\ustick{\rA}\qw&\multigate{1}{\tU}&\ustick{\rA'}\qw&\qw\\
		&\ustick{\rB}\qw&\qw&\qw&\ghost{\tU}&\ustick{\rB'}\qw&\qw	}
	\end{aligned}
	&\ =\ 
		\begin{aligned}
	\Qcircuit @C=1em @R=1em
	{	&\ustick{\rE}\qw&\qw&\qw&\multigate{1}{\tA'}&\ustick{\rE}\qw&\qw\\
		&\ustick{\rA}\qw&\multigate{1}{\tU}&\ustick{\rA'}\qw&\ghost{\tA'}&\ustick{\rA'}\qw&\qw\\
		&\ustick{\rB}\qw&\ghost{\tU}&\qw&\qw&\ustick{\rB'}\qw&\qw
	}
	\end{aligned}\ .
\label{eq:nocieq}
\end{align}
for some $\tA'\in\Trnset{\rA'\rE\to\rA'\rE}$.

Now, let us consider the special case where $\rE=\rI$, and
\begin{align}
	\begin{aligned}
	\Qcircuit @C=1em @R=1em
	{&\ustick{\rA}\qw&\gate{\tA}&\ustick{\rA}\qw&\qw}
	\end{aligned}\ =\ 
	\begin{aligned}
	\Qcircuit @C=1em @R=1em
	{&\ustick{\rA}\qw&\measureD{e}&\prepareC{\rho}&\ustick{\rA}\qw&\qw}
	\end{aligned}\ ,
\end{align}
then discard system $\rA'$ on both sides of Eq.~\eqref{eq:nocieq}. We obtain
\begin{align}
	&\begin{aligned}
	\Qcircuit @C=1em @R=1em
	{	&\ustick{\rA}\qw&\measureD{e}&\prepareC{\rho}&\ustick{\rA}\qw&\multigate{1}{\tU}&\ustick{\rA'}\qw&\measureD{e}\\
		&\ustick{\rB}\qw&\qw&\qw&\qw&\ghost{\tU}&\ustick{\rB'}\qw&\qw	}
	\end{aligned}
	\nonumber\\
	\nonumber\\
	& =\ 
		\begin{aligned}
	\Qcircuit @C=1em @R=1em
	{	&\ustick{\rA}\qw&\multigate{1}{\tU}&\ustick{\rA'}\qw&\gate{\tA'}&\ustick{\rA'}\qw&\measureD{e}\\
		&\ustick{\rB}\qw&\ghost{\tU}&\qw&\qw&\ustick{\rB'}\qw&\qw
	}
	\end{aligned}\ .
\label{eq:nosiproof1}
\end{align}
Now, since $\rho\in\aTset{\rA}_1$ is deterministic, also $\tA'\in\Trnset{\rA'\to\rA'}_1$ is deterministic, and thus
\begin{align*}
	\begin{aligned}
	\Qcircuit @C=1em @R=1em
	{	&\ustick{\rA'}\qw&\gate{\tA'}&\ustick{\rA'}\qw&\measureD{e}}
	\end{aligned}
	&\ =\ 
		\begin{aligned}
	\Qcircuit @C=1em @R=1em
	{	&\ustick{\rA'}\qw&\measureD{e}
	}
	\end{aligned}\ .
\end{align*}
Then, Eq.~\eqref{eq:nosiproof1} becomes
\begin{align}
	&\begin{aligned}
	\Qcircuit @C=1em @R=1em
	{	&\ustick{\rA}\qw&\measureD{e}&&\\
		&\ustick{\rB}\qw&\gate{\tC}&\ustick{\rB'}\qw&\qw	}
	\end{aligned}
	 \ =\ 
		\begin{aligned}
	\Qcircuit @C=1em @R=1em
	{	&\ustick{\rA}\qw&\multigate{1}{\tU}&\ustick{\rA'}\qw&\measureD{e}\\
		&\ustick{\rB}\qw&\ghost{\tU}&\ustick{\rB'}\qw&\qw
	}
	\end{aligned}\ ,
\label{eq:nosiproof2}
\end{align}
where
\begin{align}
		\begin{aligned}
	\Qcircuit @C=1em @R=1em
	{	&\ustick{\rB}\qw&\gate{\tC}&\ustick{\rB'}\qw&\qw}
	\end{aligned}\ \coloneqq\ 
	\begin{aligned}
	\Qcircuit @C=1em @R=1em
	{	\prepareC{\rho}&\ustick{\rA}\qw&\multigate{1}{\tU}&\ustick{\rA'}\qw&\measureD{e}\\
		&\ustick{\rB}\qw&\ghost{\tU}&\ustick{\rB'}\qw&\qw	}
	\end{aligned}\ .
\end{align}

The question one may ask now is whether the two conditions $\rA\not\rightarrow_\tU\rB'$ and $\rA\not\leadsto_\tU\rB'$ are equivalent. The answer is 
negative, and surprisingly the  counterexample comes from the simplest theory we have: classical theory. In classical theory system types are in 
correspondence with natural numbers $n$ representing the number of distinct pure states of the system, e.g.~for a bit pure states are $\psi_0$ and 
$\psi_1$, thus $n=2$. Parallel composition is simply given by the product rule: if $\rA=m$ and $\rB=n$, then $\rA\rB=mn$. It will turn useful to 
introduce the symbols $\mn\coloneqq\{0,1,\ldots,n-1\}$. For the sake of simplicity, given $x\in\mn$, we will 
write $x$ to denote $\psi_x\in\aTset{n}$. Channels $\tC\in\Trnset{n\to m}$ are convex combinations of maps of the form
\begin{align}
	\begin{aligned}
	\Qcircuit @C=1em @R=1em
	{	&\ustick{m}\qw&\gate{\tC}&\ustick{n}\qw&\qw}
	\end{aligned}\ =\ \sum_{x\in\mm}
			\begin{aligned}
	\Qcircuit @C=1em @R=1em
	{	&\ustick{m}\qw&\measureD{x}&\prepareC{f(x)}&\ustick{n}\qw&\qw}
	\end{aligned}\ ,
\label{eq:extclchan}
\end{align}
where $f:\mn\to\mm$ is any function. In other words, atomic transformations $\tA\in\Trnset{m\to n}$ are of the form
\begin{align*}
\begin{aligned}
	\Qcircuit @C=1em @R=1em
	{	&\ustick{m}\qw&\gate{\tA_{i,j}}&\ustick{n}\qw&\qw}
	\end{aligned}\ =\ 
	\begin{aligned}
	\Qcircuit @C=1em @R=1em
	{	&\ustick{m}\qw&\measureD{i}&\prepareC{j}&\ustick{n}\qw&\qw}
	\end{aligned}\ ,
\end{align*}
with $i\in\mm$ and $j\in\mn$.
Now, let us consider the C-not reversible map acting on a pair of bits, defined by the function  
$f(a,b)=(a\oplus b,b)$, with $\oplus$ denoting 
sum modulo 2. Notice that $\tK^2=\tI$. One has 
\begin{align}
&\begin{aligned}
	\Qcircuit @C=1em @R=1em
	{	&\ustick{\rA}\qw&\multigate{1}{\tK}&\ustick{\rA'}\qw&\measureD{e}\\
		&\ustick{\rB}\qw&\ghost{\tK}&\ustick{\rB'}\qw&\qw}
	\end{aligned}\nonumber\\
	\nonumber\\
	&=\ 
	\sum_{x,y,z=0}^1
	\begin{aligned}
	\Qcircuit @C=1em @R=1em
	{	&\ustick{\rA}\qw&\measureD{x}&\prepareC{x\oplus y}&\ustick{\rA'}\qw&\measureD{z}\\
		&\ustick{\rB}\qw&\measureD{y}&\prepareC{y}&\ustick{\rB'}\qw&\qw}
	\end{aligned}\nonumber\\
	\nonumber\\
	&=\ 
	\sum_{x,y=0}^1
	\begin{aligned}
	\Qcircuit @C=1em @R=1em
	{	&\ustick{\rA}\qw&\measureD{x}&&&\\
		&\ustick{\rB}\qw&\measureD{y}&\prepareC{y}&\ustick{\rB'}\qw&\qw}
	\end{aligned}\nonumber\\
	\nonumber\\
	&=\ 
	\begin{aligned}
	\Qcircuit @C=1em @R=1em
	{	&\ustick{\rA}\qw&\measureD{e}&&&\\
		&\ustick{\rB}\qw&\qw&\ustick{\rB'}\qw&\qw}
	\end{aligned}\ ,
\end{align}
thus $\rA\not\leadsto_\tK\rB'$. On the other hand, considering $\tC\in\Trnset{2\to2}_1$ defined by the constant function $f(x)=0$, one has
\begin{align}
&\begin{aligned}
	\Qcircuit @C=1em @R=1em
	{	&\ustick{\rA'}\qw&\multigate{1}{\tK}&\ustick{\rA}\qw&\gate{\tC}&\ustick{\rA}\qw&\multigate{1}{\tK}&\ustick{\rA'}\qw&\qw\\
		&\ustick{\rB'}\qw&\ghost{\tK}&\qw&\ustick{\rB}\qw&\qw&\ghost{\tK}&\ustick{\rB'}\qw&\qw}
	\end{aligned}\ =\ 
	\begin{aligned}
	\Qcircuit @C=1em @R=1em
	{	&\ustick{\rA}\qw&\multigate{1}{\tC'}&\ustick{\rA'}\qw&\qw\\
		&\ustick{\rB}\qw&\ghost{\tC'}&\ustick{\rB'}\qw&\qw}
	\end{aligned}\ ,
\end{align}
where $\tC'\in\Trnset{4\to4}_1$ is defined by the function $g(x,y)=(y,y)$. It is then clear that 
condition~\eqref{eq:noci} is violated, and thus
$\rA\rightarrow_\tK\rB'$. Indeed, in classical theory it is possible to copy information without disturbing, and 
thus it is also possible to interact with a system without disturbing. The C-not map $\tK$ perfectly illustrates 
this situation: it is possible to extract information from $\rB$ and leak it to $\rA'$ without affecting the 
state of system $\rB$ (this is indeed the meaning of the no-signalling condition). However, our definition 
of causal influence also accounts for the presence of a non-disturbing interaction. Looking at 
condition~\eqref{eq:nocieq}, the interpretation of the causal influence relation is clear: negating Eq.~\eqref{eq:nocieq} amounts to state that in order to replicate the effect of an intervention at 
$\rA$ after the evolution through $\tU$ one has to interact with system $\rB'$. The difference between no-signalling and no causal influence is thus
hidden, as far as we understand it from the example of classical theory, in the possibility of interacting without disturbing. We will come back on 
this observation in section~\ref{sec:niwd}, and show that this intuition can be partially turned into a general theorem.

\section{Conditions for no causal influence}\label{sec:conds}

In the present section we prove a relevant equivalent condition for $\rA\not\rightarrow_\tU\rB'$, and then 
discuss some necessary conditions. 
For our purposes it is useful to introduce the following definition
\begin{definition}
Let $\tU\in\Trnset{\rA\rB\to\rA'\rB'}_1$ be a reversible transformation. We define the reversible channel 
$\tTUA\in\Trnset{\rA_1\rA'\rB'\to\rA_1\rA'\rB'}$, where $\rA_1\simeq\rA$, as follows
\begin{align}
&\begin{aligned}
	\Qcircuit @C=1em @R=1em
	{	&\ustick{\rA_1}\qw&\multigate{2}{\tTUA}&\ustick{\rA_1}\qw&\qw\\
		&\ustick{\rA'}\qw&\ghost{\tTUA}&\ustick{\rA'}\qw&\qw\\
		&\ustick{\rB'}\qw&\ghost{\tTUA}&\ustick{\rB'}\qw&\qw}
	\end{aligned}\ \coloneqq\ 	
	\begin{aligned}
	\resizebox{0.2\textwidth}{!}{\begin{tikzpicture}
	\begin{pgfonlayer}{nodelayer}
		\node [style=none] (0) at (-3.5, 2) {};
		\node [style=none] (1) at (-3.5, 0) {};
		\node [style=none] (2) at (-3.5, -2) {};
		\node [style=none] (3) at (-2.25, 0.5) {};
		\node [style=none] (4) at (-0.5, 0.5) {};
		\node [style=none] (5) at (-2.25, -2.5) {};
		\node [style=none] (6) at (-0.5, -2.5) {};
		\node [style=none] (7) at (0.5, 0) {};
		\node [style=none] (9) at (2, 0) {};
		\node [style=none] (11) at (3, 0.5) {};
		\node [style=none] (12) at (3, -2.5) {};
		\node [style=none] (13) at (4.5, 0.5) {};
		\node [style=none] (14) at (4.5, -2.5) {};
		\node [style=none] (15) at (5.75, 0) {};
		\node [style=none] (16) at (5.75, -2) {};
		\node [style=none] (17) at (0.5, 2) {};
		\node [style=none] (18) at (2, 2) {};
		\node [style=none] (19) at (5.75, 2) {};
		\node [style=none] (20) at (-2.25, 0) {};
		\node [style=none] (21) at (-0.5, 0) {};
		\node [style=none] (22) at (-2.25, -2) {};
		\node [style=none] (23) at (-0.5, -2) {};
		\node [style=none] (24) at (3, 0) {};
		\node [style=none] (25) at (3, -2) {};
		\node [style=none] (26) at (4.5, 0) {};
		\node [style=none] (27) at (4.5, -2) {};
		\node [style=none] (28) at (-1.25, -1) {};
		\node [style=none] (30) at (-1.25, -1) {$\tU^{-1}$};
		\node [style=none] (31) at (3.75, -1) {$\tU$};
		\node [style=none] (32) at (-3, 2.5) {};
		\node [style=none] (33) at (-3, 2.5) {$\rA_1$};
		\node [style=none] (34) at (0, 0.5) {$\rA$};
		\node [style=none] (35) at (1.25, -1.5) {$\rB$};
		\node [style=none] (36) at (5.25, 2.5) {$\rA_1$};
		\node [style=none] (37) at (5.25, 0.5) {$\rA'$};
		\node [style=none] (38) at (5.25, -1.5) {$\rB'$};
		\node [style=none] (39) at (-3, 0.5) {$\rA'$};
		\node [style=none] (40) at (2.5, 0.5) {$\rA$};
		\node [style=none] (41) at (-3, -1.5) {$\rB'$};
	\end{pgfonlayer}
	\begin{pgfonlayer}{edgelayer}
		\draw (0.center) to (17.center);
		\draw (3.center) to (4.center);
		\draw (3.center) to (5.center);
		\draw (5.center) to (6.center);
		\draw (6.center) to (4.center);
		\draw (11.center) to (12.center);
		\draw (12.center) to (14.center);
		\draw (14.center) to (13.center);
		\draw (11.center) to (13.center);
		\draw (1.center) to (20.center);
		\draw (2.center) to (22.center);
		\draw (21.center) to (7.center);
		\draw (9.center) to (24.center);
		\draw (26.center) to (15.center);
		\draw (27.center) to (16.center);
		\draw (18.center) to (19.center);
		\draw [in=180, out=15, looseness=0.50] (7.center) to (18.center);
		\draw [in=-180, out=0, looseness=0.50] (17.center) to (9.center);
		\draw (23.center) to (25.center);
	\end{pgfonlayer}
\end{tikzpicture}}
	\end{aligned}
	\label{eq:deftildes}
\end{align}
\end{definition}

We state now a very useful and powerful lemma involving the above defined transformation $\tTUA$.
\begin{lemma}\label{lem:blabla}
Let $\tU\in\Trnset{\rA\rB\to\rA'\rB'}$, and $\tA\in\Trnset{\rE\rA\to\rE\rA}$. Then the following identity holds
\begin{align}
&\begin{aligned}
	\Qcircuit @C=1em @R=1em
	{	&\ustick{\rE}\qw&\qw&\qw&\multigate{1}{\tA}&\qw&\qw&\ustick{\rE}\qw&\qw\\
		&\ustick{\rA_1}\qw&\multigate{2}{\tTUA}&\ustick{\rA_1}\qw&\ghost{\tA}&\ustick{\rA_1}\qw&\multigate{2}{\tTUA}&\ustick{\rA_1}\qw&\qw\\
		&\ustick{\rA'}\qw&\ghost{\tTUA}&\qw&\ustick{\rA'}\qw&\qw&\ghost{\tTUA}&\ustick{\rA'}\qw&\qw\\
		&\ustick{\rB'}\qw&\ghost{\tTUA}&\qw&\ustick{\rB'}\qw&\qw&\ghost{\tTUA}&\ustick{\rB'}\qw&\qw}
	\end{aligned}\nonumber\\
	\nonumber\\
	&=\ 
	\begin{aligned}
	\resizebox{0.2\textwidth}{!}{\begin{tikzpicture}
	\begin{pgfonlayer}{nodelayer}
		\node [style=none] (0) at (-3.5, 2) {};
		\node [style=none] (1) at (-3.5, 0) {};
		\node [style=none] (2) at (-3.5, -2) {};
		\node [style=none] (3) at (-2.25, 0.5) {};
		\node [style=none] (4) at (-0.5, 0.5) {};
		\node [style=none] (5) at (-2.25, -2.5) {};
		\node [style=none] (6) at (-0.5, -2.5) {};
		\node [style=none] (20) at (-2.25, 0) {};
		\node [style=none] (21) at (-0.5, 0) {};
		\node [style=none] (22) at (-2.25, -2) {};
		\node [style=none] (23) at (-0.5, -2) {};
		\node [style=none] (28) at (-1.25, -1) {};
		\node [style=none] (30) at (-1.25, -1) {$\tU^{-1}$};
		\node [style=none] (32) at (-3, 2.5) {};
		\node [style=none] (33) at (-3, 2.5) {$\rA_1$};
		\node [style=none] (34) at (0, 0.5) {$\rA$};
		\node [style=none] (35) at (1.25, -1.5) {$\rB$};
		\node [style=none] (36) at (1.25, 4.5) {$\rA_1$};
		\node [style=none] (39) at (-3, 0.5) {$\rA'$};
		\node [style=none] (41) at (-3, -1.5) {$\rB'$};
		\node [style=none] (42) at (-3.5, 4) {};
		\node [style=none] (55) at (3, 0.5) {};
		\node [style=none] (56) at (3, -2.5) {};
		\node [style=none] (57) at (4.5, 0.5) {};
		\node [style=none] (58) at (4.5, -2.5) {};
		\node [style=none] (59) at (5.75, 0) {};
		\node [style=none] (60) at (5.75, -2) {};
		\node [style=none] (70) at (4.5, 0) {};
		\node [style=none] (71) at (4.5, -2) {};
		\node [style=none] (72) at (5.25, 2.5) {$\rA_1$};
		\node [style=none] (76) at (5.25, 0.5) {$\rA'$};
		\node [style=none] (77) at (5.25, -1.5) {$\rB'$};
		\node [style=none] (78) at (0.5, 2.5) {};
		\node [style=none] (79) at (2, 2.5) {};
		\node [style=none] (80) at (0.5, -0.5) {};
		\node [style=none] (81) at (2, -0.5) {};
		\node [style=none] (83) at (5.25, 4.5) {$\rE$};
		\node [style=none] (84) at (-3, 4.5) {$\rE$};
		\node [style=none] (87) at (3.75, -1) {$\tU$};
		\node [style=none] (90) at (-2, 2) {};
		\node [style=none] (91) at (-0.5, 2) {};
		\node [style=none] (92) at (-2, 4) {};
		\node [style=none] (93) at (-0.5, 4) {};
		\node [style=none] (94) at (1.25, 1) {$\tA$};
		\node [style=none] (95) at (3, 2) {};
		\node [style=none] (96) at (4.5, 2) {};
		\node [style=none] (97) at (3, 4) {};
		\node [style=none] (98) at (4.5, 4) {};
		\node [style=none] (99) at (2.5, 2.5) {$\rE$};
		\node [style=none] (100) at (0, 2.5) {$\rE$};
		\node [style=none] (101) at (0.5, 2) {};
		\node [style=none] (102) at (2, 2) {};
		\node [style=none] (103) at (0.5, 0) {};
		\node [style=none] (104) at (2, 0) {};
		\node [style=none] (105) at (5.75, 4) {};
		\node [style=none] (106) at (5.75, 2) {};
		\node [style=none] (107) at (3, 0) {};
		\node [style=none] (108) at (3, -2) {};
		\node [style=none] (109) at (2.5, 0.5) {$\rA$};
	\end{pgfonlayer}
	\begin{pgfonlayer}{edgelayer}
		\draw (3.center) to (4.center);
		\draw (3.center) to (5.center);
		\draw (5.center) to (6.center);
		\draw (6.center) to (4.center);
		\draw (1.center) to (20.center);
		\draw (2.center) to (22.center);
		\draw (55.center) to (56.center);
		\draw (56.center) to (58.center);
		\draw (58.center) to (57.center);
		\draw (55.center) to (57.center);
		\draw (70.center) to (59.center);
		\draw (71.center) to (60.center);
		\draw (78.center) to (80.center);
		\draw (80.center) to (81.center);
		\draw (81.center) to (79.center);
		\draw (78.center) to (79.center);
		\draw [in=180, out=15, looseness=0.50] (90.center) to (93.center);
		\draw [in=-180, out=0, looseness=0.50] (92.center) to (91.center);
		\draw [in=180, out=15, looseness=0.50] (95.center) to (98.center);
		\draw [in=-180, out=0, looseness=0.50] (97.center) to (96.center);
		\draw (42.center) to (92.center);
		\draw (0.center) to (90.center);
		\draw (93.center) to (97.center);
		\draw (91.center) to (101.center);
		\draw (102.center) to (95.center);
		\draw (21.center) to (103.center);
		\draw (104.center) to (107.center);
		\draw (23.center) to (108.center);
		\draw (96.center) to (106.center);
		\draw (98.center) to (105.center);
	\end{pgfonlayer}
\end{tikzpicture}}
	\end{aligned}\ ,
\label{eq:interme}
\end{align}
\end{lemma}
The proof of the above result is provided in Appendix~\ref{app:proofoflem}. The implications of Eq.~\eqref{eq:interme} reach far 
beyond the results of the present paper. Here we exploit it to prove the following key result.

\begin{theorem}
Let $\tU\in\Trnset{\rA\rB\to\rA'\rB'}_1$ be a reversible transformation. Then $\rA\not\rightarrow_\tU\rB'$ if and only if $\tTUA$ factors as
\begin{align}
&\begin{aligned}
	\Qcircuit @C=1em @R=1em
	{	&\ustick{\rA_1}\qw&\multigate{2}{\tTUA}&\ustick{\rA_1}\qw&\qw\\
		&\ustick{\rA'}\qw&\ghost{\tTUA}&\ustick{\rA'}\qw&\qw\\
		&\ustick{\rB'}\qw&\ghost{\tTUA}&\ustick{\rB'}\qw&\qw}
	\end{aligned}\ =\ 	
	\begin{aligned}
		\Qcircuit @C=1em @R=1em
	{	&\ustick{\rA_1}\qw&\multigate{1}{\TUA}&\ustick{\rA_1}\qw&\qw\\
		&\ustick{\rA'}\qw&\ghost{\TUA}&\ustick{\rA'}\qw&\qw\\
		&\ustick{\rB'}\qw&\qw&\ustick{\rB'}\qw&\qw}
	\end{aligned}\ ,
	\label{eq:nocinec}
\end{align}
where $\TUA\in\Trnset{\rA'_1\rA'\to\rA'_1\rA'}_1$.
\end{theorem}
\begin{proof}
The condition in Eq.~\eqref{eq:nocinec} is clearly necessary, as it follows from Eq.~\eqref{eq:noci} for the special case where $\rE=\rA_1\simeq\rA$ 
and $\tA=\tS$ is the swap channel for systems $\rA_1$ and $\rA$. Let us now prove that the condition is sufficient. 
Consider the result of lemma~\ref{lem:blabla}, i.e.~Eq.~\eqref{eq:interme}.
Now, if condition~\eqref{eq:nocinec} holds, the above Eq.~\eqref{eq:interme} becomes 
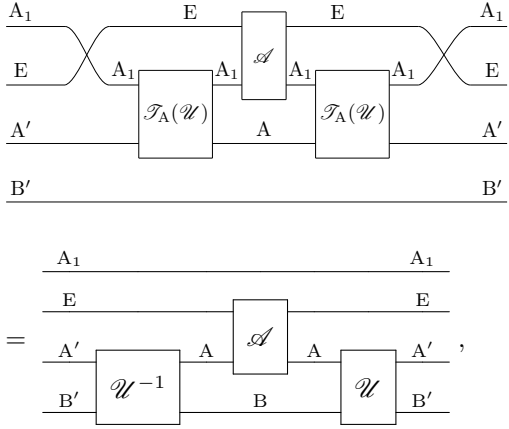
\begin{align*}
	&\begin{aligned}
	\resizebox{0.4\textwidth}{!}{\begin{tikzpicture}
	\begin{pgfonlayer}{nodelayer}
		\node [style=none] (0) at (-5, 2) {};
		\node [style=none] (1) at (-5, 0) {};
		\node [style=none] (2) at (-5, -2) {};
		\node [style=none] (33) at (-4.5, 4.5) {$\rA_1$};
		\node [style=none] (36) at (-1, 2.5) {$\rA_1$};
		\node [style=none] (39) at (-4.5, 0.5) {$\rA'$};
		\node [style=none] (41) at (-4.5, -1.5) {$\rB'$};
		\node [style=none] (42) at (-5, 4) {};
		\node [style=none] (59) at (5.5, 0) {};
		\node [style=none] (60) at (12, -2) {};
		\node [style=none] (72) at (2.5, 2.5) {$\rA_1$};
		\node [style=none] (77) at (11.5, -1.5) {$\rB'$};
		\node [style=none] (78) at (-0.5, 2.5) {};
		\node [style=none] (79) at (2, 2.5) {};
		\node [style=none] (80) at (-0.5, -0.5) {};
		\node [style=none] (81) at (2, -0.5) {};
		\node [style=none] (83) at (11.5, 2.5) {$\rE$};
		\node [style=none] (84) at (-4.5, 2.5) {$\rE$};
		\node [style=none] (90) at (-3, 2) {};
		\node [style=none] (91) at (-1.5, 2) {};
		\node [style=none] (92) at (-3, 4) {};
		\node [style=none] (93) at (-1.5, 4) {};
		\node [style=none] (95) at (9, 2) {};
		\node [style=none] (96) at (10.75, 2) {};
		\node [style=none] (97) at (9, 4) {};
		\node [style=none] (98) at (10.75, 4) {};
		\node [style=none] (100) at (1.25, 4.5) {$\rE$};
		\node [style=none] (101) at (-0.5, 2) {};
		\node [style=none] (102) at (2, 2) {};
		\node [style=none] (103) at (-0.5, 0) {};
		\node [style=none] (104) at (2, 0) {};
		\node [style=none] (105) at (12, 4) {};
		\node [style=none] (106) at (12, 2) {};
		\node [style=none] (109) at (3.75, 0.5) {$\rA$};
		\node [style=none] (110) at (11.5, 4.5) {$\rA_1$};
		\node [style=none] (111) at (3, 4.5) {};
		\node [style=none] (112) at (4.5, 4.5) {};
		\node [style=none] (113) at (3, 1.5) {};
		\node [style=none] (114) at (4.5, 1.5) {};
		\node [style=none] (115) at (3.75, 3) {$\tA$};
		\node [style=none] (116) at (3, 4) {};
		\node [style=none] (117) at (3, 2) {};
		\node [style=none] (118) at (4.5, 2) {};
		\node [style=none] (119) at (4.5, 4) {};
		\node [style=none] (121) at (0.75, 1) {$\TUA$};
		\node [style=none] (122) at (5.5, 2.5) {};
		\node [style=none] (123) at (8, 2.5) {};
		\node [style=none] (124) at (5.5, -0.5) {};
		\node [style=none] (125) at (8, -0.5) {};
		\node [style=none] (127) at (5.5, 2) {};
		\node [style=none] (128) at (12, 0) {};
		\node [style=none] (129) at (11.5, 0.5) {$\rA'$};
		\node [style=none] (130) at (8, 2) {};
		\node [style=none] (131) at (8, 0) {};
		\node [style=none] (132) at (6.25, 4.5) {$\rE$};
		\node [style=none] (133) at (5, 2.5) {$\rA_1$};
		\node [style=none] (134) at (8.5, 2.5) {$\rA_1$};
		\node [style=none] (135) at (6.75, 1) {$\TUA$};
	\end{pgfonlayer}
	\begin{pgfonlayer}{edgelayer}
		\draw (78.center) to (80.center);
		\draw (80.center) to (81.center);
		\draw (81.center) to (79.center);
		\draw (78.center) to (79.center);
		\draw [in=180, out=15, looseness=0.50] (90.center) to (93.center);
		\draw [in=-180, out=0, looseness=0.50] (92.center) to (91.center);
		\draw [in=180, out=15, looseness=0.50] (95.center) to (98.center);
		\draw [in=-180, out=0, looseness=0.50] (97.center) to (96.center);
		\draw (42.center) to (92.center);
		\draw (0.center) to (90.center);
		\draw (91.center) to (101.center);
		\draw (96.center) to (106.center);
		\draw (98.center) to (105.center);
		\draw (111.center) to (113.center);
		\draw (113.center) to (114.center);
		\draw (114.center) to (112.center);
		\draw (111.center) to (112.center);
		\draw (122.center) to (124.center);
		\draw (124.center) to (125.center);
		\draw (125.center) to (123.center);
		\draw (122.center) to (123.center);
		\draw (104.center) to (59.center);
		\draw (102.center) to (117.center);
		\draw (93.center) to (116.center);
		\draw (119.center) to (97.center);
		\draw (118.center) to (127.center);
		\draw (130.center) to (95.center);
		\draw (131.center) to (128.center);
		\draw (1.center) to (103.center);
		\draw (2.center) to (60.center);
	\end{pgfonlayer}
\end{tikzpicture}}
	\end{aligned}\\
	\\
	&=\ 
	\begin{aligned}
	\Qcircuit @C=1em @R=1em
	{	&\ustick{\rA_1}\qw&\qw&\qw&\qw&\qw&\qw&\ustick{\rA_1}\qw&\qw\\
		&\ustick{\rE}\qw&\qw&\qw&\multigate{1}{\tA}&\qw&\qw&\ustick{\rE}\qw&\qw\\
		&\ustick{\rA'}\qw&\multigate{1}{\tU^{-1}}&\ustick{\rA}\qw&\ghost{\tA}&\ustick{\rA}\qw&\multigate{1}{\tU}&\ustick{\rA'}\qw&\qw\\
		&\ustick{\rB'}\qw&\ghost{\tU^{-1}}&\qw&\ustick{\rB}\qw&\qw&\ghost{\tU}&\ustick{\rB'}\qw&\qw	}
	\end{aligned}\ ,
\end{align*}
from which one can easily conclude that
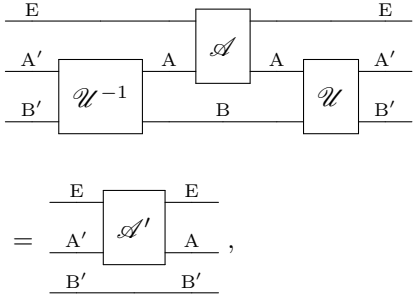
\begin{align*}
	&\begin{aligned}
	\Qcircuit @C=1em @R=1em
	{	&\ustick{\rE}\qw&\qw&\qw&\multigate{1}{\tA}&\qw&\qw&\ustick{\rE}\qw&\qw\\
		&\ustick{\rA'}\qw&\multigate{1}{\tU^{-1}}&\ustick{\rA}\qw&\ghost{\tA}&\ustick{\rA}\qw&\multigate{1}{\tU}&\ustick{\rA'}\qw&\qw\\
		&\ustick{\rB'}\qw&\ghost{\tU^{-1}}&\qw&\ustick{\rB}\qw&\qw&\ghost{\tU}&\ustick{\rB'}\qw&\qw	}
	\end{aligned}\\
	\\
	&=\ 
\begin{aligned}
	\Qcircuit @C=1em @R=1em
	{	&\ustick{\rE}\qw&\multigate{1}{\tA'}&\ustick{\rE}\qw&\qw\\
		&\ustick{\rA'}\qw&\ghost{\tA'}&\ustick{\rA}\qw&\qw\\
		&\ustick{\rB'}\qw&\qw&\ustick{\rB'}\qw&\qw	}
	\end{aligned}\ ,
\end{align*}
namely condition~\eqref{eq:noci} is satisfied.
\end{proof}

\subsection{Necessary conditions}

We already proved that no causal influence implies no-signalling. Now we provide a second necessary condition for no causal influence.
\begin{theorem}
Let $\tU\in\Trnset{\rA\rB\to\rA'\rB'}_1$ be a reversible transformation. If $\rA\not\rightarrow_\tU\rB'$, then $\tU$ can be decomposed as follows
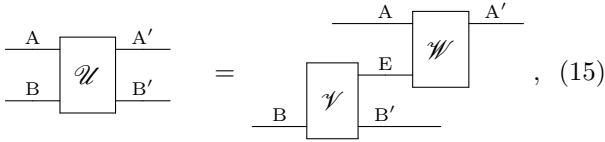
\begin{align}
	\begin{aligned}
	\Qcircuit @C=1em @R=1em
	{	&\ustick{\rA}\qw&\multigate{1}{\tU}&\ustick{\rA'}\qw&\qw\\
		&\ustick{\rB}\qw&\ghost{\tU}&\ustick{\rB'}\qw&\qw&	}
	\end{aligned}\ =\ 
\begin{aligned}
	\Qcircuit @C=1em @R=1em
	{	&&&\ustick{\rA}\qw&\multigate{1}{\tW}&\ustick{\rA'}\qw&\qw\\
		&&\pureghost{\tV}&\ustick{\rE}\qw&\ghost{\tW}&\\
		&\ustick{\rB}\qw&\multigate{-1}{\tV}&\ustick{\rB'}\qw&\qw	}
	\end{aligned}\ ,
	\label{eq:comb}
\end{align}
where $\tV\in\Trnset{\rB\to\rB'\rE}_1$ is left-reversible and $\tW\in\Trnset{\rA\rE\to\rA'}_1$ is right-reversible.
\end{theorem}
\begin{proof}
Let us consider the equivalent condition~\eqref{eq:nocinec} for no causal influence, and rewrite it in the following form
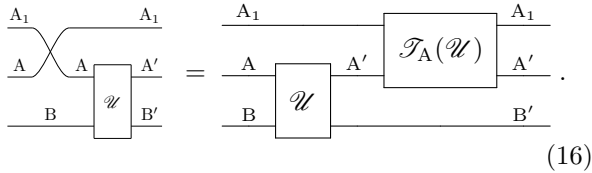
\begin{align}
	\begin{aligned}
		\resizebox{0.13\textwidth}{!}{\begin{tikzpicture}
	\begin{pgfonlayer}{nodelayer}
		\node [style=none] (0) at (-0.5, 2) {};
		\node [style=none] (7) at (0.5, 0) {};
		\node [style=none] (9) at (2, 0) {};
		\node [style=none] (11) at (3, 0.5) {};
		\node [style=none] (12) at (3, -2.5) {};
		\node [style=none] (13) at (4.5, 0.5) {};
		\node [style=none] (14) at (4.5, -2.5) {};
		\node [style=none] (15) at (5.75, 0) {};
		\node [style=none] (16) at (5.75, -2) {};
		\node [style=none] (17) at (0.5, 2) {};
		\node [style=none] (18) at (2, 2) {};
		\node [style=none] (19) at (5.75, 2) {};
		\node [style=none] (21) at (-0.5, 0) {};
		\node [style=none] (23) at (-0.5, -2) {};
		\node [style=none] (24) at (3, 0) {};
		\node [style=none] (25) at (3, -2) {};
		\node [style=none] (26) at (4.5, 0) {};
		\node [style=none] (27) at (4.5, -2) {};
		\node [style=none] (31) at (3.75, -1) {$\tU$};
		\node [style=none] (33) at (0, 2.5) {$\rA_1$};
		\node [style=none] (34) at (0, 0.5) {$\rA$};
		\node [style=none] (35) at (1.25, -1.5) {$\rB$};
		\node [style=none] (36) at (5.25, 2.5) {$\rA_1$};
		\node [style=none] (37) at (5.25, 0.5) {$\rA'$};
		\node [style=none] (38) at (5.25, -1.5) {$\rB'$};
		\node [style=none] (40) at (2.5, 0.5) {$\rA$};
	\end{pgfonlayer}
	\begin{pgfonlayer}{edgelayer}
		\draw (0.center) to (17.center);
		\draw (11.center) to (12.center);
		\draw (12.center) to (14.center);
		\draw (14.center) to (13.center);
		\draw (11.center) to (13.center);
		\draw (21.center) to (7.center);
		\draw (9.center) to (24.center);
		\draw (26.center) to (15.center);
		\draw (27.center) to (16.center);
		\draw (18.center) to (19.center);
		\draw [in=180, out=15, looseness=0.50] (7.center) to (18.center);
		\draw [in=-180, out=0, looseness=0.50] (17.center) to (9.center);
		\draw (23.center) to (25.center);
	\end{pgfonlayer}
\end{tikzpicture}}
	\end{aligned}\ =\ 
	\begin{aligned}
	\Qcircuit @C=1em @R=1em
		{	&\ustick{\rA_1}\qw&\qw&\qw&\multigate{1}{\TUA}&\ustick{\rA_1}\qw&\qw\\
			&\ustick{\rA}\qw&\multigate{1}{\tU}&\ustick{\rA'}\qw&\ghost{\TUA}&\ustick{\rA'}\qw&\qw\\
			&\ustick{\rB}\qw&\ghost{\tU}&\qw&\qw&\ustick{\rB'}\qw&\qw	}
	\end{aligned}\ .
	\label{eq:interme1}
\end{align}
Applying both sides of Eq.~\eqref{eq:interme1} to $\psi\in\aTset{\rA}_1$ and discarding $\rA_1$ on both sides by the deterministic effect $e_{\rA_1}$,
we obtain the thesis, with $\rE=\rA'$, and 
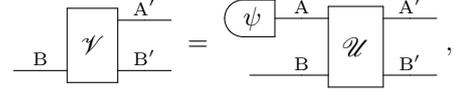
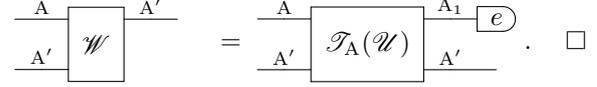
\begin{align*}
\begin{aligned}
	\Qcircuit @C=1em @R=1em
		{	&&\pureghost{\tV}&\ustick{\rA'}\qw&\qw\\
			&\ustick{\rB}\qw&\multigate{-1}{\tV}&\ustick{\rB'}\qw&\qw	}
	\end{aligned}\ =\ 
	\begin{aligned}
	\Qcircuit @C=1em @R=1em
		{	\prepareC{\psi}&\ustick{\rA}\qw&\multigate{1}{\tU}&\ustick{\rA'}\qw&\qw\\
			&\ustick{\rB}\qw&\ghost{\tU}&\ustick{\rB'}\qw&\qw	}
	\end{aligned}\ ,
\end{align*}
and
\begin{align*}
	\begin{aligned}
	\Qcircuit @C=1em @R=1em
		{	&\ustick{\rA}\qw&\multigate{1}{\tW}&\ustick{\rA'}\qw&\qw\\
			&\ustick{\rA'}\qw&\ghost{\tW}&&&}
	\end{aligned}\ =\ 
	\begin{aligned}
	\Qcircuit @C=1em @R=1em
		{	&\ustick{\rA}\qw&\multigate{1}{\TUA}&\ustick{\rA_1}\qw&\measureD{e}\\
			&\ustick{\rA'}\qw&\ghost{\TUA}&\ustick{\rA'}\qw&\qw}
	\end{aligned}\ .&&&\qedhere
\end{align*}
\end{proof}

The last result allows us to provide an alternative, simpler proof of the fact that $\rA\not\rightarrow_\tU\rB'$ implies $\rA\not\leadsto_\tU\rB'$.

\begin{theorem}
Let $\tU\in\Trnset{\rA\rB\to\rA'\rB'}_1$ be a reversible transformation. If $\rA\not\rightarrow_\tU\rB'$, then $\rA\not\leadsto_\tU\rB'$.
\end{theorem}
\begin{proof}
It is sufficient to consider the decomposition in Eq.~\eqref{eq:comb}, then discard $\rA'$ via $e_{\rA'}$. One immediately realises that condition~\eqref{eq:nosig} is satisfied, with
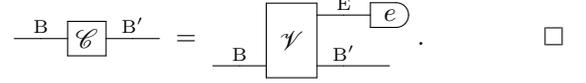
\begin{align*}
\begin{aligned}
	\Qcircuit @C=1em @R=1em
		{	&\ustick{\rB}\qw&\gate{\tC}&\ustick{\rB'}\qw&\qw}
	\end{aligned}\ =\ 
	\begin{aligned}
	\Qcircuit @C=1em @R=1em
		{	&&\pureghost{\tV}&\ustick{\rE}\qw&\measureD{e}\\
			&\ustick{\rB}\qw&\multigate{-1}{\tV}&\ustick{\rB'}\qw&\qw}
	\end{aligned}\ .&&&\qedhere
\end{align*}
\end{proof}
The latter proof highlights the following chain of implications that constitutes the general scenario in an OPT, 
given $\tU\in\Trnset{\rA\rB\to\rA'\rB'}_1$
reversible.
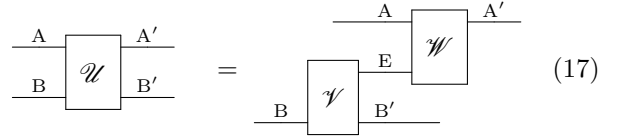
\begin{align}
\rA&\not\rightarrow_\tU\rB'\nonumber\\\nonumber\\
&\Downarrow\nonumber\\\nonumber\\
	\begin{aligned}
	\Qcircuit @C=1em @R=1em
	{	&\ustick{\rA}\qw&\multigate{1}{\tU}&\ustick{\rA'}\qw&\qw\\
		&\ustick{\rB}\qw&\ghost{\tU}&\ustick{\rB'}\qw&\qw&	}
	\end{aligned}\ &=\ 
\begin{aligned}
	\Qcircuit @C=1em @R=1em
	{	&&&\ustick{\rA}\qw&\multigate{1}{\tW}&\ustick{\rA'}\qw&\qw\\
		&&\pureghost{\tV}&\ustick{\rE}\qw&\ghost{\tW}&\\
		&\ustick{\rB}\qw&\multigate{-1}{\tV}&\ustick{\rB'}\qw&\qw	}
	\end{aligned}\label{eq:hier} \\\nonumber\\
&	\Downarrow\nonumber\\\nonumber\\
\rA&\not\leadsto_\tU\rB'\nonumber	
\end{align}
In the following sections, we will analyse cases where the above implications become equivalences.

\section{Classical theory}\label{sec:ct}

We discussed some general aspects of the notion of causal influence in classical theory in the introductory section. Now we get back to classical 
theory in the light of the results that we proved, and will show that, in classical theory, one has the following equivalence.
\begin{theorem}
Let $\tU\in\Trnset{\rA\rB\to\rA'\rB'}_1$ be reversible. If $\rA\not\leadsto_\tU\rB'$, then
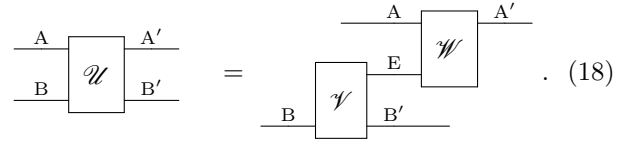
\begin{align}
	\begin{aligned}
	\Qcircuit @C=1em @R=1em
	{	&\ustick{\rA}\qw&\multigate{1}{\tU}&\ustick{\rA'}\qw&\qw\\
		&\ustick{\rB}\qw&\ghost{\tU}&\ustick{\rB'}\qw&\qw&	}
	\end{aligned}\ =\ 
\begin{aligned}
	\Qcircuit @C=1em @R=1em
	{	&&&\ustick{\rA}\qw&\multigate{1}{\tW}&\ustick{\rA'}\qw&\qw\\
		&&\pureghost{\tV}&\ustick{\rE}\qw&\ghost{\tW}&\\
		&\ustick{\rB}\qw&\multigate{-1}{\tV}&\ustick{\rB'}\qw&\qw	}
	\end{aligned}\ .
\end{align}
\end{theorem}
\begin{proof}
Let $\tU$ be reversible. The transformation $\tU$ is defined, as from Eq.~\eqref{eq:extclchan}, by two functions $f(x,y)$ and $h(x,y)$, where $x\in\mm$ 
and $y\in\mn$, provided that $\rA=m$ and $\rB=n$. Let also $\rA'=l$. Let us suppose that condition~\eqref{eq:nosig} holds. Then
\begin{align*}
&\sum_{x\in\mm,y\in\mn,z\in\ml}\	
	\begin{aligned}
	\Qcircuit @C=1em @R=1em
	{	&\ustick{\rA}\qw&\measureD{x}&\prepareC{f(x,y)}&\ustick{\rA'}\qw&\measureD{z}\\
		&\ustick{\rB}\qw&\measureD{y}&\prepareC{h(x,y)}&\ustick{\rB'}\qw&\qw	}
	\end{aligned}\\
	\\
	 =\ 
	&\sum_{x\in\mm,y\in\mn,g}p(g)\	
	\begin{aligned}
	\Qcircuit @C=1em @R=1em
	{	&\ustick{\rA}\qw&\measureD{x}&&&\\
		&\ustick{\rB}\qw&\measureD{y}&\prepareC{g(y)}&\ustick{\rB'}\qw&\qw	}
	\end{aligned}\ .
\end{align*}
This implies that, for every $x,y$, and for every $g$ such that $p(g)>0$, one has $g(y)=h(x,y)$. This implies 
that $p(g)=\delta_{g,g_0}$, and $h(x,y)=g_0(y)$. Thus
\begin{align*}
	&\begin{aligned}
	\Qcircuit @C=1em @R=1em
		{	&\ustick{\rA}\qw&\multigate{1}{\tU}&\ustick{\rA'}\qw&\qw\\
			&\ustick{\rB}\qw&\ghost{\tU}&\ustick{\rB'}\qw&\qw&	}
	\end{aligned}\ =\ \sum_{\substack{x\in\mm\\y\in\mn}}\ 
	\begin{aligned}
		\Qcircuit @C=1em @R=1em
		{	&\ustick{\rA}\qw&\measureD{x}&\prepareC{f(x,y)}&\ustick{\rA'}\qw&\qw\\
			&\ustick{\rB}\qw&\measureD{y}&\prepareC{g_0(y)}&\ustick{\rB'}\qw&\qw	}
	\end{aligned}\ ,
\end{align*}
and it is now easy to verify that condition~\eqref{eq:comb} is satisfied with
\begin{align*}
	&\begin{aligned}
	\Qcircuit @C=1em @R=1em
		{	&&\pureghost{\tV}&\ustick{\rB_1'}\qw&\qw\\
			&\ustick{\rB}\qw&\multigate{-1}{\tV}&\ustick{\rB'}\qw&\qw&	}
	\end{aligned}\ =\ 
	\begin{aligned}
	\Qcircuit @C=1em @R=1em
		{	\prepareC{0}&\ustick{\rB_1}\qw&\ghost{\tK}&\ustick{\rB_1'}\qw&\qw\\
			&\ustick{\rB}\qw&\multigate{-1}{\tK}&\ustick{\rB''}\qw&\gate{\tG}&\ustick{\rB'}\qw&\qw	}
	\end{aligned}\ ,
\end{align*}
where
\begin{align*}
	\begin{aligned}
	\Qcircuit @C=1em @R=1em
		{	&\ustick{\rB''}\qw&\gate{\tG}&\ustick{\rB'}\qw&\qw}
	\end{aligned}\ \coloneqq\sum_{y\in\mn} 
	\begin{aligned}
	\Qcircuit @C=1em @R=1em
		{	&\ustick{\rB''}\qw&\measureD{y}&\prepareC{g_0(y)}&\ustick{\rB'}\qw&\qw}
	\end{aligned}\ ,
\end{align*}
and
\begin{align*}
	&\begin{aligned}
	\Qcircuit @C=1em @R=1em
		{	&\ustick{\rA}\qw&\ghost{\tW}&\ustick{\rA'}\qw&\qw\\
			&\ustick{\rB_1}\qw&\multigate{-1}{\tW}&&&	}
	\end{aligned}\ =\ 
	\begin{aligned}
	\Qcircuit @C=1em @R=1em
		{	&\ustick{\rA}\qw&\ghost{\tU}&\ustick{\rA'}\qw&\qw\\
			&\ustick{\rB_1}\qw&\multigate{-1}{\tU}&\ustick{\rB'}\qw&\measureD{e}&	}
	\end{aligned}\ .&&\qedhere
\end{align*}
\end{proof}
As a consequence, while condition~\eqref{eq:comb} is equivalent to $\rA\not\leadsto_\tU\rB'$, there are reversible transformations $\tU$ such that
the same condition is satisfied while $\rA\rightarrow_\tU\rB'$.

\section{Quantum and Fermionic theory}\label{sec:niwd}

In the case of Quantum Theory (QT) and Fermionic Theory (FT), it is possible to prove that the hierarchy of three conditions of Eq.~\eqref{eq:hier} 
collapses in three equivalent conditions. That Eq.~\eqref{eq:comb} is equivalent to $\rA\not\leadsto_\tU\rB'$ has been proved with various techniques 
in the literature~\cite{Beckman:2001aa,Schumacher:2005aa,Eggeling_2002,Chiribella_2008,Lorenz2021causalcompositional}. We will now prove that 
condition $\rA\not\leadsto_\tU\rB'$ implies $\rA\not\rightarrow_\tU\rB'$, which is the only missing link to close the circle of implications. 
The proof uses existence and uniqueness of purification, along with existence of a pure faithful state for every system. We remind here the definition
of a faithful state for convenience of the reader.

\begin{definition}
Let $\rA\in\syst(\Theta)$, and $\Psi\in\aTset{\rA\rA'}_1$ a pure state. We say that $\Psi$ is {\em faithful} for $\rA$ if the following mapping 
is injective
\begin{align}
	\begin{aligned}
	\Qcircuit @C=1em @R=1em
		{	&\ustick{\rA}\qw&\gate{\tA}&\ustick{\rC}\qw&\qw}
	\end{aligned}\ \mapsto\ 
	\begin{aligned}
	\Qcircuit @C=1em @R=1em
		{	\multiprepareC{1}{\Psi}&\ustick{\rA}\qw&\gate{\tA}&\ustick{\rC}\qw&\qw\\
			\pureghost{\Psi}&\qw&\qw&\ustick{\rA'}\qw&\qw	}
	\end{aligned}\ ,
\end{align}
is injective.
\end{definition}
Let now $\Theta$ be a theory with unique purification and with a pure faithful state for every system $\rA$. Let also the theory be such that parallel 
composition of pure states is pure---a requirement often called {\em atomicity of parallel state composition}. Under these hypotheses one can prove 
that every test can be dilated to a reversible interaction of the system with an environment in a pure state, followed by an observation-test on the 
environment. The proof is not reported here, and can be found in~\cite{PhysRevA.81.062348,DAriano:2017aa}, along with the proof of the uniqueness results. The precise statement follows.

\begin{lemma}
Let $\Theta$ be a theory with essentially unique purification, atomicity of parallel state composition, and with a pure 
faithful state for every system $\rA$. 
Let $\tC$ be a channel $\tC\in\Trnset{\rA\to\rC}_1$. Then there exist systems $\rB,\rD$, a reversible channel $\tU\in\Trnset{\rA\rB\to\rC\rD}_1$, and a pure state $\eta\in\aTset{\rB}_1$, such that
\begin{align}
	\begin{aligned}
	\Qcircuit @C=1em @R=1em
		{	&\ustick{\rA}\qw&\gate{\tC}&\ustick{\rC}\qw&\qw}
	\end{aligned}\ =\ 
	\begin{aligned}
	\Qcircuit @C=1em @R=1em
		{	&\ustick{\rA}\qw&\multigate{1}{\tU}&\ustick{\rC}\qw&\qw\\
			\prepareC{\eta}&\ustick{\rB}\qw&\ghost{\tU}&\ustick{\rD}\qw&\measureD{e}	}
	\end{aligned}\ .
\label{eq:revdilcha}
\end{align}
We call the quadruple $(\rB,\rD,\eta,\tU)$ a {\em reversible dilation} 
of $\tC$.
\label{lem:revdil}
\end{lemma}
One can prove further results about uniqueness of reversible dilations modulo reversible transformations on $\rD$ (see Ref.~\cite{PhysRevA.81.062348}),
however we will consider here only the result discussed in the following subsection, that will turn useful in the remainder, 
and that highlights the main difference between CT discussed above on one hand, and QT and FT on the other hand.

\subsection{No interaction without disturbance}

We start proving a theorem that holds in Quantum Theory, in Real Quantum Theoory, in Fermionic Theory, as well as in any theory satisfying
the hypotheses.
\begin{theorem}[No interaction without disturbance]
Let a theory $\Theta$ satisfy the same hypotheses as lemma~\ref{lem:revdil}.
If a transformation $\tC\in\Trnset{\rA\rB\to\rC\rB}_1$ satisfies
\begin{align}
	\begin{aligned}
	\Qcircuit @C=1em @R=1em
		{	&\ustick{\rA}\qw&\multigate{1}{\tC}&\ustick{\rC}\qw&\measureD{e}\\
			&\ustick{\rB}\qw&\ghost{\tC}&\ustick{\rB}\qw&\qw	}
	\end{aligned}\ =\ 
		\begin{aligned}
	\Qcircuit @C=1em @R=1em
		{	&\ustick{\rA}\qw&\measureD{e}\\
			&\qw&\ustick{\rB}\qw&\qw&	}
	\end{aligned}\ ,
	\label{eq:non-disturbing}
\end{align}
there must exist a channel $\tD\in\Trnset{\rA\to\rC}_1$ such that
\begin{align}
	\begin{aligned}
	\Qcircuit @C=1em @R=1em
		{	&\ustick{\rA}\qw&\multigate{1}{\tC}&\ustick{\rC}\qw&\qw\\
			&\ustick{\rB}\qw&\ghost{\tC}&\ustick{\rB}\qw&\qw	}
	\end{aligned}\ =\ 
		\begin{aligned}
	\Qcircuit @C=1em @R=1em
		{	&\ustick{\rA}\qw&\gate{\tD}&\ustick{\rC}\qw&\qw\\
			&\qw&\qw&\ustick{\rB}\qw&\qw	}
	\end{aligned}\ .
	\label{eq:no-interaction}
\end{align}
\label{th:niwd}
\end{theorem}
\begin{proof}
The proof is rather straightforward, and invokes lemma~\ref{lem:revdil}, the existence of a pure faithful state, as well as essential uniqueness of 
purification. First of all, by lemma~\ref{lem:revdil} there exists $(\rE,\rF,\eta,\tU)$ with $\eta\in\aTset{\rE}_1$ pure and 
$\tU\in\Trnset{\rE\rA\rB\to\rF\rC\rB}_1$ reversible such that 
\begin{align*}
	\begin{aligned}
	\Qcircuit @C=1em @R=1em
		{	&\ustick{\rA}\qw&\multigate{1}{\tC}&\ustick{\rC}\qw&\qw\\
			&\ustick{\rB}\qw&\ghost{\tC}&\ustick{\rB}\qw&\qw	}
	\end{aligned}\ =\ 
		\begin{aligned}
	\Qcircuit @C=1em @R=1em
		{	\prepareC{\eta}&\ustick{\rE}\qw&\multigate{2}{\tU}&\ustick{\rF}\qw&\measureD{e}\\
			&\ustick{\rA}\qw&\ghost{\tU}&\ustick{\rC}\qw&\qw\\
			&\ustick{\rB}\qw&\ghost{\tU}&\ustick{\rB}\qw&\qw	}
	\end{aligned}\ .
\end{align*}
Let 
$\Phi\in\aTset{\rA\rB(\rA\rB)'}_1$ be a pure faithful state for $\rA\rB$. Then condition~\eqref{eq:non-disturbing} implies
\begin{align*}
\begin{aligned}
	\Qcircuit @C=1em @R=1em
		{	\prepareC{\eta}&\ustick{\rE}\qw&\multigate{2}{\tU}&\ustick{\rF}\qw&\measureD{e}\\
			\multiprepareC{2}{\Phi}&\ustick{\rA}\qw&\ghost{\tU}&\ustick{\rC}\qw&\measureD{e}\\
			\pureghost{\Phi}&\ustick{\rB}\qw&\ghost{\tU}&\ustick{\rB}\qw&\qw\\
			\pureghost{\Phi}&\qw&\qw&\ustick{(\rA\rB)'}\qw&\qw	}
	\end{aligned}\ =\ 
		\begin{aligned}
	\Qcircuit @C=1em @R=1em
		{	\prepareC{\eta}&\qw&\ustick{\rE}\qw&\measureD{e}\\
			\multiprepareC{2}{\Phi}&\qw&\ustick{\rA}\qw&\measureD{e}\\
			\pureghost{\Phi}&\qw&\ustick{\rB}\qw&\qw\\
			\pureghost{\Phi}&\qw&\ustick{(\rA\rB)'}\qw&\qw&	}
	\end{aligned}\ .
\end{align*}
By essential uniqueness of purification there must exist $\tV\in\Trnset{\rE\rA\to\rF\rC}_1$ such that
\begin{align*}
\begin{aligned}
	\Qcircuit @C=1em @R=1em
		{	\prepareC{\eta}&\ustick{\rE}\qw&\multigate{2}{\tU}&\ustick{\rF}\qw&\qw\\
			\multiprepareC{2}{\Phi}&\ustick{\rA}\qw&\ghost{\tU}&\ustick{\rC}\qw&\qw\\
			\pureghost{\Phi}&\ustick{\rB}\qw&\ghost{\tU}&\ustick{\rB}\qw&\qw\\
			\pureghost{\Phi}&\qw&\qw&\ustick{(\rA\rB)'}\qw&\qw	}
	\end{aligned}\ =\ 
		\begin{aligned}
	\Qcircuit @C=1em @R=1em
		{	\prepareC{\eta}&\ustick{\rE}\qw&\multigate{1}{\tV}&\ustick{\rF}\qw&\qw\\
			\multiprepareC{2}{\Phi}&\ustick{\rA}\qw&\ghost{\tV}&\ustick{\rC}\qw&\qw\\
			\pureghost{\Phi}&\qw&\qw&\ustick{\rB}\qw&\qw\\
			\pureghost{\Phi}&\qw&\qw&\ustick{(\rA\rB)'}\qw&\qw&	}
	\end{aligned}\ .
\end{align*}
As $\Phi$ is faithful, we conclude that Eq.~\eqref{eq:no-interaction} holds with
\begin{align*}
&\begin{aligned}
	\Qcircuit @C=1em @R=1em
		{	&\ustick{\rA}\qw&\gate{\tD}&\ustick{\rC}\qw&\qw}
	\end{aligned}
\ =\ 
\begin{aligned}
	\Qcircuit @C=1em @R=1em
		{	\prepareC{\eta}&\ustick{\rE}\qw&\multigate{1}{\tV}&\ustick{\rF}\qw&\measureD{e}\\
			&\ustick{\rA}\qw&\ghost{\tV}&\ustick{\rC}\qw&\qw}
	\end{aligned}\ .&&\qedhere
\end{align*}
\end{proof}
The above result holds under the hypotheses specified in the statement. However, it can be taken as a principle, that we call {\em no interacrtion 
without disturbance}. It holds as a theorem in QT, RQT, and FT. However, the results that we prove in the following hold under the general condition 
of no-interaction without disturbance, that can be precisely stated as the requirement that the above theorem~\ref{th:niwd} holds.
\begin{definition}[No interaction without disturbance]
We say that a theory $\Theta$ satisfies {\em no interaction without disturbance} if for every pair of composite systems $\rA\rB$ and $\rC\rB$, 
and for every channel $\tC\in\Trnset{\rA\rB\to\rC\rB}$, one has that, if the condition in Eq.~\eqref{eq:non-disturbing} holds, then 
the channel $\tC$ has the form given in Eq.~\eqref{eq:no-interaction}.
\label{def:niwd}
\end{definition}

Before proving the collapse of conditions, we need a further lemma.
\begin{lemma}
Let $\tU\in\Trnset{\rA\rB\to\rA'\rB'}_1$ be reversible, and $\rA\not\leadsto_\tU\rB'$. Then
\begin{align}
	\begin{aligned}
	\Qcircuit @C=1em @R=1em
		{	&\ustick{\rA'}\qw&\multigate{1}{\tU^{-1}}&\ustick{\rA}\qw&\measureD{e}\\
			&\ustick{\rB'}\qw&\ghost{\tU^{-1}}&\ustick{\rB}\qw&\gate{\tC}&\ustick{\rB'}\qw&\qw	}
	\end{aligned}\ =\ 
	\begin{aligned}
	\Qcircuit @C=1em @R=1em
		{	&\ustick{\rA'}\qw&\measureD{e}\\
			&\qw&\ustick{\rB'}\qw&\qw	}
	\end{aligned}\ ,
\end{align}
where
\begin{align*}
\begin{aligned}
	\Qcircuit @C=1em @R=1em
		{	&\ustick{\rB}\qw&\gate{\tC}&\ustick{\rB'}\qw&\qw}
	\end{aligned}\ =\ 
	\begin{aligned}
	\Qcircuit @C=1em @R=1em
		{	\prepareC{\eta}&\ustick{\rA}\qw&\multigate{1}{\tU}&\ustick{\rA'}\qw&\measureD{e}\\
			&\ustick{\rB}\qw&\ghost{\tU}&\ustick{\rB'}\qw&\qw	}
	\end{aligned}\ .
\end{align*}
\label{lem:nosiginv}
\end{lemma}
\begin{proof}
The thesis follows straightforwardly after applying $e_{\rA'}$ on both sides of the identity
\begin{align*}
	\begin{aligned}
	\Qcircuit @C=1em @R=1em
		{	&\ustick{\rA'}\qw&\multigate{1}{\tU^{-1}}&\ustick{\rA}\qw&\multigate{1}{\tU}&\ustick{\rA'}\qw&\qw\\
			&\ustick{\rB'}\qw&\ghost{\tU^{-1}}&\ustick{\rB}\qw&\ghost{\tU}&\ustick{\rB'}\qw&\qw	}
	\end{aligned}\ =\ 
	\begin{aligned}
	\Qcircuit @C=1em @R=1em
		{	&\ustick{\rA'}\qw&\qw\\
			\\
			&\ustick{\rB'}\qw&\qw	}
	\end{aligned}\ .&&&\qedhere
\end{align*}
\end{proof}

We are now ready to prove the main result of the present section.
\begin{theorem}
Let $\Theta$ satisfy no interaction without disturbance.
Let $\tU\in\Trnset{\rA\rB\to\rA'\rB'}_1$ be reversible. If $\rA\not\leadsto_\tU\rB'$, then $\rA\not\rightarrow_\tU\rB'$.
\end{theorem}
\begin{proof}
Let us consider the following identity, following from the hypothesis of no-signalling from $\rA$ to $\rB'$
\begin{align}
&\begin{aligned}
	\resizebox{0.2\textwidth}{!}{\begin{tikzpicture}
	\begin{pgfonlayer}{nodelayer}
		\node [style=none] (0) at (-3.5, 2) {};
		\node [style=none] (1) at (-3.5, 0) {};
		\node [style=none] (2) at (-3.5, -2) {};
		\node [style=none] (3) at (-2.25, 0.5) {};
		\node [style=none] (4) at (-0.5, 0.5) {};
		\node [style=none] (5) at (-2.25, -2.5) {};
		\node [style=none] (6) at (-0.5, -2.5) {};
		\node [style=none] (7) at (0.5, 0) {};
		\node [style=none] (9) at (2, 0) {};
		\node [style=none] (11) at (3, 0.5) {};
		\node [style=none] (12) at (3, -2.5) {};
		\node [style=none] (13) at (4.5, 0.5) {};
		\node [style=none] (14) at (4.5, -2.5) {};
		\node [style=none] (15) at (5.75, 0) {};
		\node [style=none] (16) at (5.75, -2) {};
		\node [style=none] (17) at (0.5, 2) {};
		\node [style=none] (18) at (2, 2) {};
		\node [style=none] (19) at (5.75, 2) {};
		\node [style=none] (20) at (-2.25, 0) {};
		\node [style=none] (21) at (-0.5, 0) {};
		\node [style=none] (22) at (-2.25, -2) {};
		\node [style=none] (23) at (-0.5, -2) {};
		\node [style=none] (24) at (3, 0) {};
		\node [style=none] (25) at (3, -2) {};
		\node [style=none] (26) at (4.5, 0) {};
		\node [style=none] (27) at (4.5, -2) {};
		\node [style=none] (28) at (-1.25, -1) {};
		\node [style=none] (30) at (-1.25, -1) {$\tU^{-1}$};
		\node [style=none] (31) at (3.75, -1) {$\tU$};
		\node [style=none] (32) at (-3, 2.5) {};
		\node [style=none] (33) at (-3, 2.5) {$\rA_1$};
		\node [style=none] (34) at (0, 0.5) {$\rA$};
		\node [style=none] (35) at (1.25, -1.5) {$\rB$};
		\node [style=none] (36) at (5.25, 2.5) {$\rA_1$};
		\node [style=none] (37) at (5.25, 0.5) {$\rA'$};
		\node [style=none] (38) at (5.25, -1.5) {$\rB'$};
		\node [style=none] (39) at (-3, 0.5) {$\rA'$};
		\node [style=none] (40) at (2.5, 0.5) {$\rA$};
		\node [style=none] (41) at (-3, -1.5) {$\rB'$};
		\node [style=none] (42) at (5.75, 2.5) {};
		\node [style=none] (43) at (5.75, 1.5) {};
		\node [style=none] (44) at (6.75, 2) {};
		\node [style=none] (45) at (5.75, 0.5) {};
		\node [style=none] (46) at (5.75, -0.5) {};
		\node [style=none] (47) at (6.75, 0) {};
		\node [style=none] (48) at (6.25, 2) {$e$};
		\node [style=none] (49) at (6.25, 0) {$e$};
	\end{pgfonlayer}
	\begin{pgfonlayer}{edgelayer}
		\draw (0.center) to (17.center);
		\draw (3.center) to (4.center);
		\draw (3.center) to (5.center);
		\draw (5.center) to (6.center);
		\draw (6.center) to (4.center);
		\draw (11.center) to (12.center);
		\draw (12.center) to (14.center);
		\draw (14.center) to (13.center);
		\draw (11.center) to (13.center);
		\draw (1.center) to (20.center);
		\draw (2.center) to (22.center);
		\draw (21.center) to (7.center);
		\draw (9.center) to (24.center);
		\draw (26.center) to (15.center);
		\draw (27.center) to (16.center);
		\draw (18.center) to (19.center);
		\draw [in=180, out=15, looseness=0.50] (7.center) to (18.center);
		\draw [in=-180, out=0, looseness=0.50] (17.center) to (9.center);
		\draw (23.center) to (25.center);
		\draw (42.center) to (43.center);
		\draw [in=-90, out=0, looseness=1.25] (43.center) to (44.center);
		\draw [in=90, out=0, looseness=1.25] (42.center) to (44.center);
		\draw (45.center) to (46.center);
		\draw [in=-90, out=0, looseness=1.25] (46.center) to (47.center);
		\draw [in=90, out=0, looseness=1.25] (45.center) to (47.center);
	\end{pgfonlayer}
\end{tikzpicture}}
\end{aligned}\ 
	=\ 
	\begin{aligned}
	\Qcircuit @C=1em @R=1em
		{	&\ustick{\rA_1}\qw &\qw                    &\qw            &\measureD{e}\\
			&\ustick{\rA'}\qw&\multigate{1}{\tU^{-1}}&\ustick{\rA}\qw&\measureD{e}\\
			&\ustick{\rB'}\qw&\ghost{\tU^{-1}}&\ustick{\rB}\qw&\gate{\tC}&\ustick{\rB'}\qw&\qw	}
	\end{aligned}\ .
\label{eq:beff}
\end{align}
Now, considering that by virtue of lemma~\ref{lem:nosiginv} one has
\begin{align*}
	\begin{aligned}
	\Qcircuit @C=1em @R=1em
		{	&\ustick{\rA_1}\qw &\qw                    &\qw            &\measureD{e}\\
			&\ustick{\rA'}\qw&\multigate{1}{\tU^{-1}}&\ustick{\rA}\qw&\measureD{e}\\
			&\ustick{\rB'}\qw&\ghost{\tU^{-1}}&\ustick{\rB}\qw&\gate{\tC}&\ustick{\rB'}\qw&\qw	}
	\end{aligned}\ =\ 
	\begin{aligned}
	\Qcircuit @C=1em @R=1em
		{	&\ustick{\rA_1}\qw &\measureD{e}\\
			&\ustick{\rA'}\qw&\measureD{e}\\
			&\qw&\ustick{\rB'}\qw&\qw	}
	\end{aligned}\ ,
\end{align*}
we can invoke definition~\ref{th:niwd} to conclude that there must exist $\TUA\in\Trnset{\rA_1\rA'\to\rA_1\rA'}$ such that
\begin{align*}
	\begin{aligned}
	\resizebox{0.2\textwidth}{!}{\begin{tikzpicture}
	\begin{pgfonlayer}{nodelayer}
		\node [style=none] (0) at (-3.5, 2) {};
		\node [style=none] (1) at (-3.5, 0) {};
		\node [style=none] (2) at (-3.5, -2) {};
		\node [style=none] (3) at (-2.25, 0.5) {};
		\node [style=none] (4) at (-0.5, 0.5) {};
		\node [style=none] (5) at (-2.25, -2.5) {};
		\node [style=none] (6) at (-0.5, -2.5) {};
		\node [style=none] (7) at (0.5, 0) {};
		\node [style=none] (9) at (2, 0) {};
		\node [style=none] (11) at (3, 0.5) {};
		\node [style=none] (12) at (3, -2.5) {};
		\node [style=none] (13) at (4.5, 0.5) {};
		\node [style=none] (14) at (4.5, -2.5) {};
		\node [style=none] (15) at (5.75, 0) {};
		\node [style=none] (16) at (5.75, -2) {};
		\node [style=none] (17) at (0.5, 2) {};
		\node [style=none] (18) at (2, 2) {};
		\node [style=none] (19) at (5.75, 2) {};
		\node [style=none] (20) at (-2.25, 0) {};
		\node [style=none] (21) at (-0.5, 0) {};
		\node [style=none] (22) at (-2.25, -2) {};
		\node [style=none] (23) at (-0.5, -2) {};
		\node [style=none] (24) at (3, 0) {};
		\node [style=none] (25) at (3, -2) {};
		\node [style=none] (26) at (4.5, 0) {};
		\node [style=none] (27) at (4.5, -2) {};
		\node [style=none] (28) at (-1.25, -1) {};
		\node [style=none] (30) at (-1.25, -1) {$\tU^{-1}$};
		\node [style=none] (31) at (3.75, -1) {$\tU$};
		\node [style=none] (32) at (-3, 2.5) {};
		\node [style=none] (33) at (-3, 2.5) {$\rA_1$};
		\node [style=none] (34) at (0, 0.5) {$\rA$};
		\node [style=none] (35) at (1.25, -1.5) {$\rB$};
		\node [style=none] (36) at (5.25, 2.5) {$\rA_1$};
		\node [style=none] (37) at (5.25, 0.5) {$\rA'$};
		\node [style=none] (38) at (5.25, -1.5) {$\rB'$};
		\node [style=none] (39) at (-3, 0.5) {$\rA'$};
		\node [style=none] (40) at (2.5, 0.5) {$\rA$};
		\node [style=none] (41) at (-3, -1.5) {$\rB'$};
	\end{pgfonlayer}
	\begin{pgfonlayer}{edgelayer}
		\draw (0.center) to (17.center);
		\draw (3.center) to (4.center);
		\draw (3.center) to (5.center);
		\draw (5.center) to (6.center);
		\draw (6.center) to (4.center);
		\draw (11.center) to (12.center);
		\draw (12.center) to (14.center);
		\draw (14.center) to (13.center);
		\draw (11.center) to (13.center);
		\draw (1.center) to (20.center);
		\draw (2.center) to (22.center);
		\draw (21.center) to (7.center);
		\draw (9.center) to (24.center);
		\draw (26.center) to (15.center);
		\draw (27.center) to (16.center);
		\draw (18.center) to (19.center);
		\draw [in=180, out=15, looseness=0.50] (7.center) to (18.center);
		\draw [in=-180, out=0, looseness=0.50] (17.center) to (9.center);
		\draw (23.center) to (25.center);
	\end{pgfonlayer}
\end{tikzpicture}}
	\end{aligned}
	\ =\ 
	\begin{aligned}
		\Qcircuit @C=1em @R=1em
	{	&\ustick{\rA_1}\qw&\multigate{1}{\TUA}&\ustick{\rA_1}\qw&\qw\\
		&\ustick{\rA'}\qw&\ghost{\TUA}&\ustick{\rA'}\qw&\qw\\
		&\ustick{\rB'}\qw&\qw&\ustick{\rB'}\qw&\qw}
	\end{aligned}\ .
\end{align*}
The latter is precisely condition~\eqref{eq:nocinec}.
\end{proof}
In conclusion, we then proved that in a theory where no interaction without disturbance holds, the two conditions of no signalling and no causal 
influence are equivalent. This is then true, in particular, in theories where the purification property holds.
\subsection{Interaction without disturbance}
What can we say in the case of violation of no interaction without disturbance? In order to analyse this question, it is useful to prove the following
result.
\begin{theorem}
\label{th:nocinodistfactid}
Let $\rA\not\to_\tU\rB$, and suppose that
\begin{align}
	\begin{aligned}
	\Qcircuit @C=1em @R=1em
		{	&\ustick{\rA}\qw&\multigate{1}{\tU}&\ustick{\rA}\qw&\measureD{e}\\
			&\ustick{\rB}\qw&\ghost{\tU}&\ustick{\rB}\qw&\qw	}
	\end{aligned}\ =\ 
		\begin{aligned}
	\Qcircuit @C=1em @R=1em
		{	&\ustick{\rA}\qw&\measureD{e}\\
			&\qw&\ustick{\rB}\qw&\qw&	}
	\end{aligned}\ .
	\label{eq:norevint}
\end{align}
Then
\begin{align}
	\begin{aligned}
	\Qcircuit @C=1em @R=1em
		{	&\ustick{\rA}\qw&\multigate{1}{\tU}&\ustick{\rA}\qw&\qw\\
			&\ustick{\rB}\qw&\ghost{\tU}&\ustick{\rB}\qw&\qw	}
	\end{aligned}\ =\ 
		\begin{aligned}
	\Qcircuit @C=1em @R=1em
		{	&\ustick{\rA}\qw&\gate{\tV}&\ustick{\rA}\qw&\qw\\
			&\qw&\ustick{\rB}\qw&\qw&\qw}
	\end{aligned}\ .
	\label{eq:factorid}
\end{align}
\end{theorem}
\begin{proof}
The hypothesis that $\rA\not\to_\tU\rB$ can be expressed equivalently by Eq.~\eqref{eq:interme1}. Combining the latter with Eq.~\eqref{eq:norevint}, 
we have
\begin{align*}
	\begin{aligned}
		\resizebox{0.1\textwidth}{!}{\begin{tikzpicture}
	\begin{pgfonlayer}{nodelayer}
		\node [style=none] (0) at (-0.5, 2) {};
		\node [style=none] (7) at (0.5, 0) {};
		\node [style=none] (9) at (2, 0) {};
		\node [style=none] (17) at (0.5, 2) {};
		\node [style=none] (18) at (2, 2) {};
		\node [style=none] (19) at (3, 2) {};
		\node [style=none] (21) at (-0.5, 0) {};
		\node [style=none] (23) at (-0.5, -2) {};
		\node [style=none] (24) at (3, 0) {};
		\node [style=none] (25) at (3, -2) {};
		\node [style=none] (33) at (0, 2.5) {$\rA_1$};
		\node [style=none] (34) at (0, 0.5) {$\rA$};
		\node [style=none] (35) at (1.25, -1.5) {$\rB$};
		\node [style=none] (36) at (2.5, 2.5) {$\rA_1$};
		\node [style=none] (40) at (2.5, 0.5) {$\rA$};
		\node [style=none] (41) at (3, 0.5) {};
		\node [style=none] (42) at (3, -0.5) {};
		\node [style=none] (43) at (4, 0) {};
		\node [style=none] (44) at (3.5, 0) {$e$};
	\end{pgfonlayer}
	\begin{pgfonlayer}{edgelayer}
		\draw (0.center) to (17.center);
		\draw (21.center) to (7.center);
		\draw (9.center) to (24.center);
		\draw (18.center) to (19.center);
		\draw [in=180, out=15, looseness=0.50] (7.center) to (18.center);
		\draw [in=-180, out=0, looseness=0.50] (17.center) to (9.center);
		\draw (23.center) to (25.center);
		\draw [in=90, out=0] (41.center) to (43.center);
		\draw (42.center) to (41.center);
		\draw [in=-90, out=0] (42.center) to (43.center);
	\end{pgfonlayer}
\end{tikzpicture}}
	\end{aligned}\ =\ 
	\begin{aligned}
	\Qcircuit @C=1em @R=1em
		{	&\ustick{\rA_1}\qw&\qw&\qw&\multigate{1}{\TUA}&\ustick{\rA_1}\qw&\qw\\
			&\ustick{\rA}\qw&\multigate{1}{\tU}&\ustick{\rA}\qw&\ghost{\TUA}&\ustick{\rA}\qw&\measureD{e}\\
			&\ustick{\rB}\qw&\ghost{\tU}&\qw&\qw&\ustick{\rB}\qw&\qw	}
	\end{aligned}\ ,
\end{align*}
and inverting $\tU$ on both sides, and composing with an arbitrary deterministic state of $\rA_1$, one has
\begin{align*}
	\begin{aligned}
	\Qcircuit @C=1em @R=1em
		{	&\ustick{\rA}\qw&\multigate{1}{\tU^{-1}}&\ustick{\rA_1}\qw&\qw\\
			&\ustick{\rB}\qw&\ghost{\tU^{-1}}&\ustick{\rB}\qw&\qw	}
	\end{aligned}\ =\ 
		\begin{aligned}
	\Qcircuit @C=1em @R=1em
		{	&\ustick{\rA}\qw&\gate{\tV}&\ustick{\rA_1}\qw&\qw\\
			&\qw&\ustick{\rB}\qw&\qw&\qw	}
	\end{aligned}\ ,
\end{align*}
where
\begin{align*}
		\begin{aligned}
	\Qcircuit @C=1em @R=1em
		{	&\ustick{\rA}\qw&\gate{\tV}&\ustick{\rA_1}\qw&\qw}
	\end{aligned}\ \coloneqq\ 
			\begin{aligned}
	\Qcircuit @C=1em @R=1em
		{	\prepareC{\rho}&\ustick{\rA_1}\qw&\multigate{1}{\TUA}&\ustick{\rA_1}\qw&\qw\\
			&\ustick{\rA}\qw&\ghost{\TUA}&\ustick{\rA}\qw&\measureD{e}	}
	\end{aligned}\ .
\end{align*}
By reversibility of $\tU$, one has that also $\tV$ has to be invertible. Finally, this implies the thesis.
\end{proof}

As a consequence of the above result, if a theory $\Theta$ has interactions without disturbance, and one of those---say 
$\tU\in\Trnset{\rA\rB\to\rA\rB}$---is reversible, by definition $\tU$ satisfies Eq.~\eqref{eq:norevint} but not Eq.~\eqref{eq:factorid}. Then, by
virtue of theorem~\ref{th:nocinodistfactid}, it cannot hold that $\rA\not\to_\tU\rB$. As a consequence, it is not true in $\Theta$ that 
$\rA\not\leadsto_\tV\rB$ implies $\rA\not\to_\tV\rB$.

\section{The $T$-process}\label{sec:tproc}

Let us now consider a reversible transformation $\tU\in\Trnset{\rA\to\rA'}$, where $\rA=\rA_1\rA_2\ldots\rA_N$ and $\rA'=\rA_1'\rA_2'\ldots\rA_M'$. We 
wonder how to characterise the {\em neighnourhood} of $\rA_i$, defined as the subset of the set 
$N^+_{\rA_i}\subseteq\{\rA_1',\rA_2',\ldots,\rA_M'\}$  of output systems of $\tU$, such that $\rA_i\to_\tU\rA'_j$ if and only if 
$\rA_j'\in N^+_{\rA_i}$. Following condition~\eqref{eq:nocinec}, we observe that the (reversible) channel $\TUAvar{\tU}{\rA_i}$ allows us to 
immediately identify the systems in $N^+_{\rA_i}$. Indeed, by definition of $N^+_{\rA_i}$, we have the following result
\begin{lemma}
Let $\tU\in\Trnset{\rA_1\rA_2\ldots\rA_N\to\rA_1'\rA_2'\ldots\rA_N'}$, and consider the input system $\rA_i$. Then 
\begin{align}
\tTUAvar{\tU}{\rA_i}=\TUAvar{\tU}{\rA_i}\otimes\tI_{\bar N^+_{\rA_i}}, 
\end{align}
where $\bar X$ denotes the complement of $X$ in $\{\rA_1',\rA_2',\ldots,\rA_M'\}$, and for every $\rA_j'\in N^+_{\rA_i}$ one cannot have 
\begin{align}
\TUAvar{\tU}{\rA_i}=\tI_{\rA'_j}\otimes\tV. 
\end{align}
In other words, $N^+_{\rA_i}$ is the largest subset $X$ of $\{\rA_1'\rA_2'\ldots\rA_M'\}$ such that for $\rA_j'\in X$ one has 
\begin{align}
\tTUAvar{\tU}{\rA_i}\neq\tI_{\rA'_j}\otimes\tV.
\end{align}
\label{lem:condt}
\end{lemma}
By the above arguments it is clear that $\TUAvar{\tU}{\rA_i}\in\Trnset{\rA_i N^+_{\rA_i}\to\rA_i N^+_{\rA_i}}$.
Moreover, since the swap operator $\tS_{\rA_i}$ for systems $\rA_i$ and $\tilde\rA_i\simeq\rA_i$ commutes with $\tS_{\rA_j}$ for $\rA_i\neq\rA_j$, and
conjugation by a reversible transformation $\tU\otimes\tI$ preserves commutation, one has 
\begin{align*}
\TUAvar{\tU}{\rA_i}\TUAvar{\tU}{\rA_j}=\TUAvar{\tU}{\rA_j}\TUAvar{\tU}{\rA_i}\, 
\end{align*}
where we implicitly pad the transformations $\TUAvar{\tU}{\rA_k}$ with identity transformations on suitable systems in order to make them act on the 
same system. We omit the identity transformations for the sake of a light notation.

We call the transformation $\TUAvar{\tU}{\rA_i}$ the {\em $T$-process} of $\tU$ relative to $\rA_i$. The advantage of defining the $T$-process is
that it is easily calculated through Eq.~\eqref{eq:deftildes}, 
and it contains all the information needed to determine $N^+_{\rA_i}$ through the condition given by lemma~\ref{lem:condt}. Moreover, by a calculation 
identical to that of Eq.~\eqref{eq:interme}, one can easily show that if one considers the composite system $\rA_i\rA_j$, one has 
$N^+_{\rA_i\rA_j}=N^+_{\rA_i}\cup N^+_{\rA_j}$, and 
$\TUAvar{\tU}{\rA_i\rA_j}=\TUAvar{\tU}{\rA_i}\TUAvar{\tU}{\rA_j}=\TUAvar{\tU}{\rA_j}\TUAvar{\tU}{\rA_i}$. In 
other words, the $T$-process detects all causal influence relations system-wise, and this excludes
activation of causal influence by non-local interventions.

\section{Conclusion}\label{sec:conc}

In this study we analysed the notion of causal influence in the context of operational probabilistic theories. 
More precisely, we studied under what circumstances we can say that a reversible process involving two (or more) 
input systems and two (or more) output systems allows for causal influence from one of the inputs to one of the 
outputs. We started generalising two alternate definitions taken from the literature on quantum information 
processing. In the general scenario, the two notions turned out to be different, and we adopted as a  definition 
of causal influence the weaker one, taking signalling as a strictly necessary condition for causal influence. We 
then discussed necessary  and sufficient conditions and the related definition of $T$-process of a reversible 
transformation $\tU$ relative to one of its input systems $\rA$. A special instance of $T$-process was used for 
quantum cellular automata in Ref.~\cite{ARRIGHI2011372} for the purpose of decomposing a special family of 
QCAs in local gates.

While in Quantum theory the two notions coincide, we discussed the case of classical theory, which provides an instructive example of the case where
no-signalling is not sufficient for no-causal influence. The analysis allowed us to identify an intermediate condition which is necessary for no-causal influence and sufficient for no-signalling, which can be summarised by requiring that the process has to have the structure of a channel with 
memory. While in the classical case this condition coincides with no-signalling, and is thus strictly weaker than no-causal influence, at the time 
of writing we do not have examples of theories where the no-signalling condition is strictly weaker than the memory channel structure.

We also identified a feature of a theory is sufficient for the three conditions to become equivalent: this feature was introduced as no-interaction 
without disturbance. Its failure in the case of a reversible process determines causal influence without signalling. While the relation of no 
interaction without disturbance with {\em no information without disturbance}~\cite{DAriano2020information} is beyond the scope of the present paper, 
we plan to study it in the near future.

Interestingly, the example that we spotted of a reversible process that allows for interaction without 
disturbance is the classical controlled-not.
This gate allows indeed the agent that controls the target to copy the content of the control without affecting 
it. We deem such an intervention to
have a causal influence on the control, even if it does not affect its state. Indeed, it is clear that one cannot 
copy the state of a 
system without an effective interaction. Secondarily, we claim that e.g.~copying information from a system has a 
causal influence on it, even though 
its \emph{local} state might be unaffected, as one can figure out thinking of many examples in everyday life. The
influence consists in creating correlations between the system and its environment, that were absent before 
occurrence of the interaction.

The definition of causal influence that we propose here has a consequence on the notion of neighbourhood of a cell in a cellular automaton. Indeed,
while the definition based on signalling causes puzzling questions in the case of classical cellular automata---such as seemingly local cellular 
automata whose inverse is non local---a definition based on causal influence sheds some light on the mentioned issues, showing that the actual 
neighbourhood can be much larger than the signalling neighbourhood. This phenomenon lies at the basis of the effect that was noticed in 
Ref.~\cite{PhysRevA.95.012331}, where the ``quantised'' version of a classical cellular automaton was observed to produce a ``speedup'' in the 
propagation of information. In the perspective of our definition, the effect is due to the fact that the process of ``quantising'' a classical gate 
turns its neighbourhood of causal influence---which might be strictly larger than its signalling neighbourhood---into the neighbourhood of the 
quantised version (in the quantum case there is no distinction between the two neighbourhoods). As an example, the quantum controlled-not exhibits a 
kickback on the control, highlighting that the latter is subject to a causal influence from the target.


\acknowledgments
The author acknowledges useful discussions about the topic of the present paper with Alessandro Bisio, Alessandro 
Tosini, Francesco Buscemi, Robin Lorenz, and Jon Barrett. This work was supported by ``MIUR Dipartimenti di 
Eccellenza 2018-2022 project F11I18000680001''.

\bibliography{causal-influ}
\bibliographystyle{unsrturl}

\begin{appendix}
\section{Proof of Lemma~\ref{lem:blabla}}\label{app:proofoflem}
Here we report the proof of Lemma~\ref{lem:blabla}.
\begin{proof}
By definition of $\tTUA$, one has
\begin{align}
&\begin{aligned}
	\Qcircuit @C=1em @R=1em
	{	&\ustick{\rE}\qw&\qw&\qw&\multigate{1}{\tA}&\qw&\qw&\ustick{\rE}\qw&\qw\\
		&\ustick{\rA_1}\qw&\multigate{2}{\tTUA}&\ustick{\rA_1}\qw&\ghost{\tA}&\ustick{\rA_1}\qw&\multigate{2}{\tTUA}&\ustick{\rA_1}\qw&\qw\\
		&\ustick{\rA'}\qw&\ghost{\tTUA}&\qw&\ustick{\rA'}\qw&\qw&\ghost{\tTUA}&\ustick{\rA'}\qw&\qw\\
		&\ustick{\rB'}\qw&\ghost{\tTUA}&\qw&\ustick{\rB'}\qw&\qw&\ghost{\tTUA}&\ustick{\rB'}\qw&\qw}
	\end{aligned}\nonumber\\
	\nonumber\\
	&\ =\ 
	\begin{aligned}
	\resizebox{0.35\textwidth}{!}{\begin{tikzpicture}
	\begin{pgfonlayer}{nodelayer}
		\node [style=none] (0) at (-3.5, 2) {};
		\node [style=none] (1) at (-3.5, 0) {};
		\node [style=none] (2) at (-3.5, -2) {};
		\node [style=none] (3) at (-2.25, 0.5) {};
		\node [style=none] (4) at (-0.5, 0.5) {};
		\node [style=none] (5) at (-2.25, -2.5) {};
		\node [style=none] (6) at (-0.5, -2.5) {};
		\node [style=none] (7) at (0.5, 0) {};
		\node [style=none] (9) at (2, 0) {};
		\node [style=none] (11) at (2.75, 0.5) {};
		\node [style=none] (12) at (2.75, -2.5) {};
		\node [style=none] (13) at (4.25, 0.5) {};
		\node [style=none] (14) at (4.25, -2.5) {};
		\node [style=none] (17) at (0.5, 2) {};
		\node [style=none] (18) at (2, 2) {};
		\node [style=none] (19) at (4.5, 2) {};
		\node [style=none] (20) at (-2.25, 0) {};
		\node [style=none] (21) at (-0.5, 0) {};
		\node [style=none] (22) at (-2.25, -2) {};
		\node [style=none] (23) at (-0.5, -2) {};
		\node [style=none] (24) at (2.75, 0) {};
		\node [style=none] (25) at (2.75, -2) {};
		\node [style=none] (26) at (4.25, 0) {};
		\node [style=none] (27) at (4.25, -2) {};
		\node [style=none] (28) at (-1.25, -1) {};
		\node [style=none] (30) at (-1.25, -1) {$\tU^{-1}$};
		\node [style=none] (31) at (3.5, -1) {$\tU$};
		\node [style=none] (32) at (-3, 2.5) {};
		\node [style=none] (33) at (-3, 2.5) {$\rA_1$};
		\node [style=none] (34) at (0, 0.5) {$\rA$};
		\node [style=none] (35) at (1.25, -1.5) {$\rB$};
		\node [style=none] (36) at (3.5, 2.5) {$\rA_1$};
		\node [style=none] (37) at (5.25, 0.5) {$\rA'$};
		\node [style=none] (38) at (5.25, -1.5) {$\rB'$};
		\node [style=none] (39) at (-3, 0.5) {$\rA'$};
		\node [style=none] (40) at (2.25, 0.5) {$\rA$};
		\node [style=none] (41) at (-3, -1.5) {$\rB'$};
		\node [style=none] (42) at (-3.5, 4) {};
		\node [style=none] (43) at (4.5, 4) {};
		\node [style=none] (44) at (6, 4) {};
		\node [style=none] (46) at (6, 2) {};
		\node [style=none] (49) at (6.25, 0.5) {};
		\node [style=none] (50) at (8, 0.5) {};
		\node [style=none] (51) at (6.25, -2.5) {};
		\node [style=none] (52) at (8, -2.5) {};
		\node [style=none] (53) at (9, 0) {};
		\node [style=none] (54) at (10.5, 0) {};
		\node [style=none] (55) at (11.25, 0.5) {};
		\node [style=none] (56) at (11.25, -2.5) {};
		\node [style=none] (57) at (12.75, 0.5) {};
		\node [style=none] (58) at (12.75, -2.5) {};
		\node [style=none] (59) at (14, 0) {};
		\node [style=none] (60) at (14, -2) {};
		\node [style=none] (61) at (9, 2) {};
		\node [style=none] (62) at (10.5, 2) {};
		\node [style=none] (63) at (14, 2) {};
		\node [style=none] (64) at (6.25, 0) {};
		\node [style=none] (65) at (8, 0) {};
		\node [style=none] (66) at (6.25, -2) {};
		\node [style=none] (67) at (8, -2) {};
		\node [style=none] (68) at (11.25, 0) {};
		\node [style=none] (69) at (11.25, -2) {};
		\node [style=none] (70) at (12.75, 0) {};
		\node [style=none] (71) at (12.75, -2) {};
		\node [style=none] (72) at (7, 2.5) {$\rA_1$};
		\node [style=none] (73) at (8.5, 0.5) {$\rA$};
		\node [style=none] (74) at (9.75, -1.5) {$\rB$};
		\node [style=none] (75) at (13.5, 2.5) {$\rA_1$};
		\node [style=none] (76) at (13.5, 0.5) {$\rA'$};
		\node [style=none] (77) at (13.5, -1.5) {$\rB'$};
		\node [style=none] (78) at (4.5, 4.5) {};
		\node [style=none] (79) at (6, 4.5) {};
		\node [style=none] (80) at (4.5, 1.5) {};
		\node [style=none] (81) at (6, 1.5) {};
		\node [style=none] (82) at (14, 4) {};
		\node [style=none] (83) at (13.5, 4.5) {$\rE$};
		\node [style=none] (84) at (-3, 4.5) {$\rE$};
		\node [style=none] (85) at (5.25, 3) {$\tA$};
		\node [style=none] (86) at (7.25, -1) {$\tU^{-1}$};
		\node [style=none] (87) at (12, -1) {$\tU$};
		\node [style=none] (88) at (10.75, 0.5) {$\rA$};
	\end{pgfonlayer}
	\begin{pgfonlayer}{edgelayer}
		\draw (0.center) to (17.center);
		\draw (3.center) to (4.center);
		\draw (3.center) to (5.center);
		\draw (5.center) to (6.center);
		\draw (6.center) to (4.center);
		\draw (11.center) to (12.center);
		\draw (12.center) to (14.center);
		\draw (14.center) to (13.center);
		\draw (11.center) to (13.center);
		\draw (1.center) to (20.center);
		\draw (2.center) to (22.center);
		\draw (21.center) to (7.center);
		\draw (9.center) to (24.center);
		\draw (18.center) to (19.center);
		\draw [in=180, out=15, looseness=0.50] (7.center) to (18.center);
		\draw [in=-180, out=0, looseness=0.50] (17.center) to (9.center);
		\draw (23.center) to (25.center);
		\draw (46.center) to (61.center);
		\draw (49.center) to (50.center);
		\draw (49.center) to (51.center);
		\draw (51.center) to (52.center);
		\draw (52.center) to (50.center);
		\draw (55.center) to (56.center);
		\draw (56.center) to (58.center);
		\draw (58.center) to (57.center);
		\draw (55.center) to (57.center);
		\draw (65.center) to (53.center);
		\draw (54.center) to (68.center);
		\draw (70.center) to (59.center);
		\draw (71.center) to (60.center);
		\draw (62.center) to (63.center);
		\draw [in=180, out=15, looseness=0.50] (53.center) to (62.center);
		\draw [in=-180, out=0, looseness=0.50] (61.center) to (54.center);
		\draw (67.center) to (69.center);
		\draw (26.center) to (64.center);
		\draw (27.center) to (66.center);
		\draw (42.center) to (43.center);
		\draw (44.center) to (82.center);
		\draw (78.center) to (80.center);
		\draw (80.center) to (81.center);
		\draw (81.center) to (79.center);
		\draw (78.center) to (79.center);
	\end{pgfonlayer}
\end{tikzpicture}}
	\end{aligned}\nonumber\\
	\nonumber\\
	&=\ 
	\begin{aligned}
	\resizebox{0.3\textwidth}{!}{\begin{tikzpicture}
	\begin{pgfonlayer}{nodelayer}
		\node [style=none] (0) at (-3.5, 2) {};
		\node [style=none] (1) at (-3.5, 0) {};
		\node [style=none] (2) at (-3.5, -2) {};
		\node [style=none] (3) at (-2.25, 0.5) {};
		\node [style=none] (4) at (-0.5, 0.5) {};
		\node [style=none] (5) at (-2.25, -2.5) {};
		\node [style=none] (6) at (-0.5, -2.5) {};
		\node [style=none] (7) at (0.5, 0) {};
		\node [style=none] (9) at (2, 0) {};
		\node [style=none] (17) at (0.5, 2) {};
		\node [style=none] (18) at (2, 2) {};
		\node [style=none] (19) at (3, 2) {};
		\node [style=none] (20) at (-2.25, 0) {};
		\node [style=none] (21) at (-0.5, 0) {};
		\node [style=none] (22) at (-2.25, -2) {};
		\node [style=none] (23) at (-0.5, -2) {};
		\node [style=none] (24) at (5.5, 0) {};
		\node [style=none] (25) at (8, -2) {};
		\node [style=none] (28) at (-1.25, -1) {};
		\node [style=none] (30) at (-1.25, -1) {$\tU^{-1}$};
		\node [style=none] (32) at (-3, 2.5) {};
		\node [style=none] (33) at (-3, 2.5) {$\rA_1$};
		\node [style=none] (34) at (0, 0.5) {$\rA$};
		\node [style=none] (35) at (3.75, -1.5) {$\rB$};
		\node [style=none] (36) at (2.5, 2.5) {$\rA_1$};
		\node [style=none] (39) at (-3, 0.5) {$\rA'$};
		\node [style=none] (40) at (3.75, 0.5) {$\rA$};
		\node [style=none] (41) at (-3, -1.5) {$\rB'$};
		\node [style=none] (42) at (-3.5, 4) {};
		\node [style=none] (43) at (3, 4) {};
		\node [style=none] (46) at (4.5, 2) {};
		\node [style=none] (53) at (5.5, 0) {};
		\node [style=none] (54) at (7, 0) {};
		\node [style=none] (55) at (8, 0.5) {};
		\node [style=none] (56) at (8, -2.5) {};
		\node [style=none] (57) at (9.5, 0.5) {};
		\node [style=none] (58) at (9.5, -2.5) {};
		\node [style=none] (59) at (10.75, 0) {};
		\node [style=none] (60) at (10.75, -2) {};
		\node [style=none] (61) at (5.5, 2) {};
		\node [style=none] (62) at (7, 2) {};
		\node [style=none] (63) at (10.75, 2) {};
		\node [style=none] (68) at (8, 0) {};
		\node [style=none] (69) at (8, -2) {};
		\node [style=none] (70) at (9.5, 0) {};
		\node [style=none] (71) at (9.5, -2) {};
		\node [style=none] (72) at (7.75, 2.5) {$\rA_1$};
		\node [style=none] (75) at (10.25, 2.5) {$\rA_1$};
		\node [style=none] (76) at (10.25, 0.5) {$\rA'$};
		\node [style=none] (77) at (10.25, -1.5) {$\rB'$};
		\node [style=none] (78) at (3, 4.5) {};
		\node [style=none] (79) at (4.5, 4.5) {};
		\node [style=none] (80) at (3, 1.5) {};
		\node [style=none] (81) at (4.5, 1.5) {};
		\node [style=none] (83) at (10.5, 4.5) {$\rE$};
		\node [style=none] (84) at (-3, 4.5) {$\rE$};
		\node [style=none] (85) at (3.75, 3) {$\tA$};
		\node [style=none] (87) at (8.75, -1) {$\tU$};
		\node [style=none] (88) at (4.5, 4) {};
		\node [style=none] (89) at (11, 4) {};
	\end{pgfonlayer}
	\begin{pgfonlayer}{edgelayer}
		\draw (0.center) to (17.center);
		\draw (3.center) to (4.center);
		\draw (3.center) to (5.center);
		\draw (5.center) to (6.center);
		\draw (6.center) to (4.center);
		\draw (1.center) to (20.center);
		\draw (2.center) to (22.center);
		\draw (21.center) to (7.center);
		\draw (9.center) to (24.center);
		\draw (18.center) to (19.center);
		\draw [in=180, out=15, looseness=0.50] (7.center) to (18.center);
		\draw [in=-180, out=0, looseness=0.50] (17.center) to (9.center);
		\draw (23.center) to (25.center);
		\draw (46.center) to (61.center);
		\draw (55.center) to (56.center);
		\draw (56.center) to (58.center);
		\draw (58.center) to (57.center);
		\draw (55.center) to (57.center);
		\draw (54.center) to (68.center);
		\draw (70.center) to (59.center);
		\draw (71.center) to (60.center);
		\draw (62.center) to (63.center);
		\draw [in=180, out=15, looseness=0.50] (53.center) to (62.center);
		\draw [in=-180, out=0, looseness=0.50] (61.center) to (54.center);
		\draw (42.center) to (43.center);
		\draw (78.center) to (80.center);
		\draw (80.center) to (81.center);
		\draw (81.center) to (79.center);
		\draw (78.center) to (79.center);
		\draw (88.center) to (89.center);
	\end{pgfonlayer}
\end{tikzpicture}}
	\end{aligned}\nonumber\\
	\nonumber\\
	&=\ 
	\begin{aligned}
	\resizebox{0.2\textwidth}{!}{\begin{tikzpicture}
	\begin{pgfonlayer}{nodelayer}
		\node [style=none] (0) at (-3.5, 2) {};
		\node [style=none] (1) at (-3.5, 0) {};
		\node [style=none] (2) at (-3.5, -2) {};
		\node [style=none] (3) at (-2.25, 0.5) {};
		\node [style=none] (4) at (-0.5, 0.5) {};
		\node [style=none] (5) at (-2.25, -2.5) {};
		\node [style=none] (6) at (-0.5, -2.5) {};
		\node [style=none] (20) at (-2.25, 0) {};
		\node [style=none] (21) at (-0.5, 0) {};
		\node [style=none] (22) at (-2.25, -2) {};
		\node [style=none] (23) at (-0.5, -2) {};
		\node [style=none] (28) at (-1.25, -1) {};
		\node [style=none] (30) at (-1.25, -1) {$\tU^{-1}$};
		\node [style=none] (32) at (-3, 2.5) {};
		\node [style=none] (33) at (-3, 2.5) {$\rA_1$};
		\node [style=none] (34) at (0, 0.5) {$\rA$};
		\node [style=none] (35) at (1.25, -1.5) {$\rB$};
		\node [style=none] (36) at (1.25, 4.5) {$\rA_1$};
		\node [style=none] (39) at (-3, 0.5) {$\rA'$};
		\node [style=none] (41) at (-3, -1.5) {$\rB'$};
		\node [style=none] (42) at (-3.5, 4) {};
		\node [style=none] (55) at (3, 0.5) {};
		\node [style=none] (56) at (3, -2.5) {};
		\node [style=none] (57) at (4.5, 0.5) {};
		\node [style=none] (58) at (4.5, -2.5) {};
		\node [style=none] (59) at (5.75, 0) {};
		\node [style=none] (60) at (5.75, -2) {};
		\node [style=none] (70) at (4.5, 0) {};
		\node [style=none] (71) at (4.5, -2) {};
		\node [style=none] (72) at (5.25, 2.5) {$\rA_1$};
		\node [style=none] (76) at (5.25, 0.5) {$\rA'$};
		\node [style=none] (77) at (5.25, -1.5) {$\rB'$};
		\node [style=none] (78) at (0.5, 2.5) {};
		\node [style=none] (79) at (2, 2.5) {};
		\node [style=none] (80) at (0.5, -0.5) {};
		\node [style=none] (81) at (2, -0.5) {};
		\node [style=none] (83) at (5.25, 4.5) {$\rE$};
		\node [style=none] (84) at (-3, 4.5) {$\rE$};
		\node [style=none] (87) at (3.75, -1) {$\tU$};
		\node [style=none] (90) at (-2, 2) {};
		\node [style=none] (91) at (-0.5, 2) {};
		\node [style=none] (92) at (-2, 4) {};
		\node [style=none] (93) at (-0.5, 4) {};
		\node [style=none] (94) at (1.25, 1) {$\tA$};
		\node [style=none] (95) at (3, 2) {};
		\node [style=none] (96) at (4.5, 2) {};
		\node [style=none] (97) at (3, 4) {};
		\node [style=none] (98) at (4.5, 4) {};
		\node [style=none] (99) at (2.5, 2.5) {$\rE$};
		\node [style=none] (100) at (0, 2.5) {$\rE$};
		\node [style=none] (101) at (0.5, 2) {};
		\node [style=none] (102) at (2, 2) {};
		\node [style=none] (103) at (0.5, 0) {};
		\node [style=none] (104) at (2, 0) {};
		\node [style=none] (105) at (5.75, 4) {};
		\node [style=none] (106) at (5.75, 2) {};
		\node [style=none] (107) at (3, 0) {};
		\node [style=none] (108) at (3, -2) {};
		\node [style=none] (109) at (2.5, 0.5) {$\rA$};
	\end{pgfonlayer}
	\begin{pgfonlayer}{edgelayer}
		\draw (3.center) to (4.center);
		\draw (3.center) to (5.center);
		\draw (5.center) to (6.center);
		\draw (6.center) to (4.center);
		\draw (1.center) to (20.center);
		\draw (2.center) to (22.center);
		\draw (55.center) to (56.center);
		\draw (56.center) to (58.center);
		\draw (58.center) to (57.center);
		\draw (55.center) to (57.center);
		\draw (70.center) to (59.center);
		\draw (71.center) to (60.center);
		\draw (78.center) to (80.center);
		\draw (80.center) to (81.center);
		\draw (81.center) to (79.center);
		\draw (78.center) to (79.center);
		\draw [in=180, out=15, looseness=0.50] (90.center) to (93.center);
		\draw [in=-180, out=0, looseness=0.50] (92.center) to (91.center);
		\draw [in=180, out=15, looseness=0.50] (95.center) to (98.center);
		\draw [in=-180, out=0, looseness=0.50] (97.center) to (96.center);
		\draw (42.center) to (92.center);
		\draw (0.center) to (90.center);
		\draw (93.center) to (97.center);
		\draw (91.center) to (101.center);
		\draw (102.center) to (95.center);
		\draw (21.center) to (103.center);
		\draw (104.center) to (107.center);
		\draw (23.center) to (108.center);
		\draw (96.center) to (106.center);
		\draw (98.center) to (105.center);
	\end{pgfonlayer}
\end{tikzpicture}}
	\end{aligned}\ ,
\end{align}
where for the third equality we used property~\ref{it:braid} of parallel composition in section~\ref{sec:opts}.
\end{proof}
\end{appendix}

\end{document}